# FERROELECTRIC THIN FILMS PHASE DIAGRAMS WITH SELF-POLARIZED PHASE AND ELECTRET STATE.

M.D.Glinchuk*, A.N.Morozovska**, E.A.Eliseev*

*Institute for Problems of Materials Science, NAS of Ukraine,
Krjijanovskogo 3, 03142 Kiev, Ukraine, glin@materials.kiev.ua

**V. Lashkaryov Institute of Semiconductor Physics, NAS of Ukraine,
41, pr. Nauki, 03028 Kiev, Ukraine, morozo@mail.i.com.ua

## Abstract

In present work we calculated the three components of polarization in phenomenological theory framework by consideration of three Euler-Lagrange equations, which include mismatch effect and influence of misfit dislocations, surface piezoelectric effect caused by broken symmetry on the film surface, surface tension and depolarization field. These equations were solved with the help of variational method proposed by us earlier. This approach lead to the free energy in the form of algebraic equation of different powers of polarization components with the coefficients dependent on film thickness, mismatch effect, temperature and other parameters. Several new terms proportional to misfit strain appeared in the free energy expression: built-in electric field normal to the surface originated from piezoelectricity in vicinity of surface even for the cubic symmetry of bulk ferroelectrics, odd powers of normal to the surface component of polarization. The obtained free energy made it possible to calculate all properties of the film by conventional procedure of minimization. As an example we calculated phase diagrams of PZT 50/50 films on different substrates that lead to compressive or tensile strain. The calculations of pyroelectric coefficient and dielectric permittivity temperature dependencies had shown the electret-like polar state, e.g. existence of pyroelectricity, below the critical thickness of ferroelectric-paraelectric phase transitions. Our theory predicts, that mismatch-induced field $E_m$ could be compatible with thermodynamic coercive field and thus cause self-polarization in thin ferroelectric films.

## 1. INTRODUCTION

Ferroelectric thin films and their multilayers are widely applied in microelectronics and optoelectronics and in particular in non-volatile ferroelectric random access memories [1], [2]. The maximum miniaturization of the chips and devices based on these materials is quite urgent. Because of this fundamental questions, the thickness dependence of ferroelectric properties and the possible disappearance of ferroelectricity at a finite critical thickness become crucial. It was generally accepted, that because of ferroelectricity cooperative nature it could not could exist in the strongly restricted systems. In particular on the basis of empirical evidence (see e.g. [3]), ferroelectricity was believed to disappear in the nano-sized film with thickness $h$ below some critical value $h_{cr}$ ($h<h_{cr}$~10-100nm). However, first principle theoretical works [4], [5], and recent experimental papers demonstrate the existence of the ferroelectric properties in the films with thickness up to few monolayers [6], [7], [8], [9]. Although these first-principle simulations predict ferroelectric ground states for various $ABO_3$ thin films under the condition of vanishing internal electric field several important physical features (e.g. strained state imposed by substrate) were omitted. More realistic first-principle calculations [10] showed the disappearance of ferroelectricity at critical thickness about several *nm*. Therefore first-principle simulations lead to contradictory results, because it is cumbersome to take into account real confinement conditions including substrate, electrodes, film-substrate inter-influence.

The phenomenological description allows including the contribution of all these components of the real heterostructure into free energy functional (see e.g. [11], [12], [13], [14]). In the paper [14] we had shown the existence of internal electric field caused by misfit strain between the film and its substrate and piezoelectric coupling that exists in the vicinity of the surface. This normal to the film surface built-in electric field was supposed to be the physical reason for the phenomenon of film self-polarization (see [15], [16], [17]) and polarization conservation in a few monolayers film. Note, that recently [18] surface was considered as random field defect, so that electric field originated from the surface was included into free energy functional. However all the calculations in [14] were carried out on the basis of renormalized free energy with only one component of spontaneous polarization $P_Z$ normal to the surface. Because three components of polarization $P_X$, $P_Y$ and $P_Z$ coupled with one another usually exist in perovskite family of ferroelectrics, all of them should be considered.

In the present paper we carried out such calculations allowing for misfit strain, originated from the difference between lattice constrains and thermal expansion coefficients of the film and substrate, misfit dislocations, which tend to decrease the strain, depolarization field for case of superconducting electrodes and surface energy related to surface tension. The calculations were performed for the case of monodomain film of insulator ferroelectric epitaxially grown on the substrate and superconducting electrodes. This type of film and electrodes was used earlier in phenomenological calculations [19] and recently in first principle calculations [10]. Three coupled Euler-Lagrange equations for $P_X$, $P_Y$, $P_Z$ were solved with the help of variational method, that allows us to obtain free energy of polynomial type, but with renormalized coefficients that depend on film thickness and other characteristics of heterostructure. The appearance of the odd powers of $P_Z$ related to internal built-in electric field is the peculiar feature of the free energy. This free energy allows one to calculate the properties by conventional minimization procedure. The calculation of the properties and phase diagram with special attention to the influence of built-in misfit-induced field was performed. It was shown, that in general case there is no transition from ferroelectric to paraelectric phase at film thickness $l < l_{cr}$, but rather to electret-like state (see definition in [20]) induced by this field. The conditions of appearance of electret state and self-polarized ferroelectric phase are calculated and discussed. The comparison of the theory with available experimental data for thin films showed, that the theory describes the results of measurements pretty good.

## 2. FREE ENERGY FUNCTIONAL

Let us consider ferroelectric thin film with the thickness $l$ $(-l/2 \le z \le l/2)$on the thick substrate in the external electric field $\mathbf{E}$. In phenomenological theory approach free energy functional can be written in the form:

$$\Delta G = G - G_0 = \int g_V dV + \int g_S ds . \qquad \text{(1a)}$$

Here the first and the second integral reflect polarization dependent contribution of the bulk and the surface of the film, $G_0$ is polarization independent part of the free energy. The bulk free energy density $g_V$ can be represented as:

$$g_V = g_{0V} + \frac{1}{2}\left( \delta_Z \left(\frac{dP_Z}{dz}\right)^2 + \delta_X \left(\frac{dP_X}{dz}\right)^2 + \delta_X \left(\frac{dP_Y}{dz}\right)^2 \right) - \mathbf{P}\,\mathbf{E} - P_Z \frac{E_d}{2} \qquad \text{(1b)}$$

Here the energy $g_{0V}$ represents the expansion over polarization powers with coefficients renormalized by mechanical tension via electrostriction on the base of procedure developed in [21]; $E_d$ is depolarization field, its value for the case of single-domain insulator film with super-conducting electrodes can be written in the form [19]:

$$E_d = 4\pi\left(\overline{P}_Z - P_Z(z)\right). \qquad \text{(2)}$$

Hereafter the bar over a letter denoting a physical quantity represents the spatial averaging over the film thickness, e.g. $\overline{P}_Z \equiv \frac{1}{l}\int_{-l/2}^{l/2} dz P_Z(z)$. Allowing for surface energy is related to surface tension [22], it is possible to represent the second term in Eq. (1) similarly to [14], [23]:

$$G_s = \sum_{i=1}^{2} \int \mu_i u_{xx}^{(i)} u_{yy}^{(i)} dxdy . \qquad \text{(3a)}$$

Here parameters $\mu$ and $u_{jj}$ ($j = x, y$) are respectively the surface tension coefficient and strain tensor components, $i = 1, 2$ reflects the contribution of the film two surfaces.

In what follows we will consider two main contributions to the strain tensor, proposed in [14] related to surface polarization $P_Z(\pm l/2)$ via piezoelectric effect that exists even in a cubic symmetry lattice near the film surface, and the second is related to mismatch effects discussed in the introduction. For perovskite structure lattice the only components of piezoelectric effect coefficient $d_{kjj}$, that couple strains $u_{xx}^{(i)}, u_{yy}^{(i)}$ with polarization $P_Z(\pm l/2)$ are $d_{zxx}^{(i)}, d_{zyy}^{(i)}$ [24]. Therefore

$$u_{xx}^{(i)} = u_{xxm}^{(i)} + d_{zxx}^{(i)} P_{zi}^{(i)}, \quad u_{yy}^{(i)} = u_{yym}^{(i)} + d_{zyy}^{(i)} P_{zi}^{(i)}, \tag{3b}$$

where $z_1 = l/2$, $z_2 = -l/2$, $u_{jjm}$ is the tensor of mechanical strain that is proportional to the difference of the lattice constants and thermal expansion coefficients of a substrate and a film. In considered case $u_{xx}^{(i)} = u_{yy}^{(i)}$, $d_{zxx}^{(i)} = d_{zyy}^{(i)} \equiv d_{31}$ and $u_{xxm}^{(i)} = u_{yym}^{(i)} \equiv u_m^{(i)}$ so the product $u_{xx}^{(i)} u_{yy}^{(i)}$ in Eq. (3a) can be rewritten as $u_{xx}^{(i)} u_{yy}^{(i)} = \left(u_m^{(i)} + d_{31} P_{zi}^{(i)}\right)^2$, where the term $u_m^{(i)}$ is independent on polarization, while the other terms are defined the surface free energy. In what follows we will consider the realistic situation of the film on the substrate with freestanding upper surface, where parameters $u_m^{(1)} = 0$ and $u_m^{(2)} = U_m$. For the description of the phase transitions in the films with perovskite structure the free energy (1) with respect to the Eqs.(3) acquires the form $\Delta G = \Delta G_V + \Delta G_S$:

$$\Delta G_V = \frac{1}{l}\int_{-l/2}^{l/2} dz \left[\begin{array}{l} \frac{\alpha_Z(T,U_m)}{2} P_Z^2(z) + \frac{\alpha_X(T,U_m)}{2}\left(P_X^2(z) + P_Y^2(z)\right) + \frac{\eta_X}{2} P_X^2(z)\cdot P_Y^2(z) + \\ + \frac{\eta_Z}{2} P_Z^2(z)\left(P_X^2(z) + P_Y^2(z)\right) + \frac{\beta_Z}{4} P_Z^4(z) + \frac{\gamma_Z}{6} P_Z^6(z) + \frac{\beta_X}{4}\left(P_X^4(z) + P_Y^4(z)\right) \\ + \frac{\gamma_X}{6}\left(P_X^6(z) + P_Y^6(z)\right) + \frac{\gamma_{XYZ}}{2} P_Z^2(z) P_X^2(z) P_Y^2(z) + \\ + \frac{\eta_{XYZ}}{4}\left(P_Z^4(z)\left(P_X^2(z) + P_Y^2(z)\right) + P_X^4(z)\left(P_Y^2(z) + P_Z^2(z)\right) + P_Y^4(z)\left(P_X^2(z) + P_Z^2(z)\right)\right) + \\ + \frac{1}{2}\left(\delta_Z\left(\frac{dP_Z(z)}{dz}\right)^2 + \delta_X\left(\frac{dP_X(z)}{dz}\right)^2 + \delta_X\left(\frac{dP_Y(z)}{dz}\right)^2\right) - P_Z(z)\left(E_Z + 2\pi\left(\bar{P}_Z - P_Z(z)\right)\right) - \\ - P_X(z)E_X - P_Y(z)E_Y \end{array}\right] \tag{4a}$$

$$\Delta G_S = \frac{\delta_Z}{2l}\left[\frac{P_Z^2(l/2)}{\lambda_{Z1}} + \frac{\left(P_Z(-l/2) + P_m\right)^2}{\lambda_{Z2}}\right] + \frac{\delta_X}{2\lambda_X l}\left[P_X^2(l/2) + P_X^2(-l/2) + P_Y^2(l/2) + P_Y^2(-l/2)\right] \tag{4b}$$

Hereinafter $\gamma_{Z,X} > 0$, $\gamma_{XYZ} > 0$, $\eta_{XYZ} > 0$, $\delta_{Z,X} > 0$, while $\beta_{Z,X} < 0$ (first order phase transitions) $\beta_{Z,X} > 0$ (second order phase transitions) and we introduced the extrapolation lengths $\lambda_X = \lambda_Y$, $\lambda_{Z1,2}$ and "misfit-induced" surface polarization $P_m$ in the form:

$$\lambda_{Z1,2} = \frac{\delta_Z}{2\mu_{1,2} d_{31}^2}, \qquad P_m = \frac{U_m}{d_{31}}. \tag{5}$$

Since the signs of the parameters $U_m$ and $d_{31}$ can be positive or negative both $P_m > 0$ and $P_m < 0$ are expected. The extrapolation lengths $\lambda_{X,Z1,2}$ can be only positive because $\mu_{1,2} > 0$ [22]. The renormalized by mechanical tension coefficient $\alpha$ in Eq.(4a) has the form [21]:

$$\alpha_{X,Z}(T) = \alpha_T\left(T - T_C^{X,Z}\right), \quad T_C^Z = T_C + \frac{2Q_{12}U_m^*}{\alpha_T(S_{11} + S_{12})}, \quad T_C^X = T_C + \frac{(Q_{11} + Q_{12})U_m^*}{\alpha_T(S_{11} + S_{12})}. \tag{6}$$

Here parameters $T_C$, $\alpha_T$, $Q_{11}$, $Q_{12}$ and $S_{11}$, $S_{12}$ are respectively ferroelectric transition temperature, inverse Curie constant, electrostriction coefficient and elastic modulus regarded known for the bulk material (e.g. for perovskites $Q_{12} < 0$ and $(Q_{11} + Q_{12}) > 0$, $(S_{11} + S_{12}) > 0$). In (6) we take into account, that in accordance with [25], [26] the "effective" substrate constant is renormalized by misfit dislocations appeared at critical thickness $l_d \sim 1/|U_m|$, namely:

$$U_m^*(T,l)=\frac{b^*(T,l)-a(T)}{b^*(T,l)},\qquad b^*(T,l)=\begin{cases} b(T)\left(1-U_m(T)\left(1-\frac{l_d}{l}\right)\right), & l>l_d \\ b(T), & l\le l_d \end{cases} \tag{7}$$

Temperature dependence in Eq.(7) is related to thermal expansion coefficients of the film ($\rho_a$) and substrate ($\rho_b$) so that $a(T)\approx a(T_g)\left(1+(T-T_g)\rho_a\right)$ and $b(T)\approx b(T_g)\left(1+(T-T_g)\rho_b\right)$ are film and substrate lattice constants respectively, $T_g$ is film growth temperature. However on the surface $z=-l/2$ (i.e. in the surface energy $\Delta G_S$ and thus in (5)) the real misfit strain $U_m(T)$ exists, namely:

$$U_m(T)=\frac{b(T)-a(T)}{b(T)}\approx\frac{b(T_g)-a(T_g)}{b(T_g)}+(\rho_b-\rho_a)(T-T_g) \tag{8}$$

Having substituted (8) into (7) one obtains:

$$U_m^*(T,l)=\frac{b^*(T,l)-a(T)}{b^*(T,l)}=\begin{cases} \frac{l_d}{l}\frac{U_m(T)}{1-U_m(T)(1-l_d/l)}, & l>l_d \\ U_m(T), & l\le l_d \end{cases} \tag{9}$$

In order to estimate the temperature dependence of misfit strain $U_m$, we use material coefficients for 50/50 PZT films $S_{11}=10.5\cdot10^{-12}\,m^2/N$, $S_{12}=-3.7\cdot10^{-12}\,m^2/N$, $Q_{11}=0.0966\,m^4/C^2$, $Q_{12}=-0.0460\,m^4/C^2$ [27], all other material coefficients are taken from [28]. We obtain that $|T-T_g|\le 500K$, $|\rho_b-\rho_a|\le 5\cdot10^{-6}$ and $|(\rho_b-\rho_a)(T-T_g)|\le 2.5\cdot10^{-3}$, whereas $|1-a(T_g)/b(T_g)|\ge 1.5\cdot10^{-2}$ for PZT films on typical metal-oxide substrates. So, misfit strain in (9) appeared practically independent on temperature and thus one can use approximation $U_m\approx 1-a(T_g)/b(T_g)$.

## 3. FREE ENERGY WITH RENORMALIZED COEFFICIENTS

The coupled equations for the polarization components can be obtained by variation over polarization of free energy functional (4). This yields the following Euler-Lagrange equations with the boundary conditions:

$$\begin{cases} P_Z\left(\alpha_Z+\eta_Z\left(P_X^2+P_Y^2\right)\right)+\beta_Z P_Z^3+f_Z-\delta_Z\left(\frac{d^2P_Z}{dz^2}\right)=E_Z+4\pi\left(\overline{P}_Z-P_Z\right), \\ \left.\left(P_Z+\lambda_{Z1}\frac{dP_Z}{dz}\right)\right|_{z=l/2}=0,\qquad \left.\left(P_Z-\lambda_{Z2}\frac{dP_Z}{dz}\right)\right|_{z=-l/2}=-P_m. \end{cases} \tag{10a}$$

$$\begin{cases} P_X\left(\alpha_X+\eta_Z P_Z^2+\eta_X P_Y^2\right)+\beta_X P_X^3+f_X-\delta_X\left(\frac{d^2P_X}{dz^2}\right)=E_X, \\ \left.\left(P_X+\lambda_X\frac{dP_X}{dz}\right)\right|_{z=l/2}=0,\qquad \left.\left(P_X-\lambda_X\frac{dP_X}{dz}\right)\right|_{z=-l/2}=0. \end{cases} \tag{10b}$$

$$\begin{cases} P_Y\left(\alpha_X+\eta_Z P_Z^2+\eta_X P_X^2\right)+\beta_X P_Y^3+f_Y-\delta_X\left(\dfrac{d^2P_Y}{dz^2}\right)=E_Y, \\ \left.\left(P_Y+\lambda_X\dfrac{dP_Y}{dz}\right)\right|_{z=l/2}=0, \qquad \left.\left(P_Y-\lambda_X\dfrac{dP_Y}{dz}\right)\right|_{z=-l/2}=0. \end{cases} \tag{10c}$$

Here we gathered the terms with fifth power of polarization into the following expression: $f_i=\gamma_i P_i^5+\gamma_{XYZ}P_i\,P_j^2P_k^2+\eta_{XYZ}P_i^3\left(P_j^2+P_k^2\right)+\eta_{XYZ}P_i\,\left(P_j^4+P_k^4\right)/2$, $(i,j,k=x,y,z,\ \ i\neq j\neq k)$. Note, that for the second order phase transitions terms in $f_i$ can be neglected. Let us find as it was proposed earlier in [12] the approximate solution of the nonlinear Eqs.(10) by the direct variational method. We will choose the one-parametric trial functions similarly to [14], in the form of solutions of linearized Eqs.(10) with external field components $E_Z$, $E_X$, $E_Y$, that satisfy the boundary conditions. In the case of different extrapolation lengths $\lambda_{Z1}\neq\lambda_{Z2}$ trial functions become very cumbersome (see e.g. [12], [29]). In order to demonstrate the mismatch effect influence more clearly, hereinafter we put $\lambda_{Z1}=\lambda_{Z2}=\lambda_Z$ and use the following trial functions (see Appendix A):

$$P_Z(z)=P_{VZ}\frac{1-\varphi(z)}{1-\overline{\overline{\varphi(z)}}}-\frac{P_m}{2}\left(\varphi(z)-\xi(z)\right), \qquad P_{X,Y}(z)=P_{V\,X,Y}\frac{1-\phi(z)}{1-\overline{\overline{\phi(z)}}}. \tag{11}$$

The variational parameters - amplitudes $P_{VX,Y,Z}$ must be determined by the minimization of the free energy (4). Hereinafter we used the following functions:

$$\varphi(z)=\frac{ch(z/l_Z)}{ch(l/2l_Z)+(\lambda_Z/l_Z)sh(l/2l_Z)}, \qquad \xi(z)=\frac{sh(z/l_Z)}{sh(l/2l_Z)+(\lambda_Z/l_Z)ch(l/2l_Z)}, \tag{12a}$$

$$\phi(z)=\begin{cases} \dfrac{ch(z/l_X)}{ch(l/2l_X)+(\lambda_X/l_X)sh(l/2l_X)}, & \alpha_X>0 \\ \dfrac{\cos(z/l_X)}{\cos(l/2l_X)-(\lambda_X/l_X)\sin(l/2l_X)}, & \alpha_X<0 \end{cases} \tag{12b}$$

Here $l_Z$ and $l_X$ are respectively the longitudinal and transverse characteristic lengths: $l_Z=\sqrt{\delta_Z/(4\pi+\alpha_Z)}\approx\sqrt{\delta_Z/4\pi}$ and $l_X(T,U_m)=\sqrt{\delta_X/|\alpha_X(T,U_m)|}$. For the majority of ferroelectrics $l_Z\sim 1\div 10\,\mathring{A}$ [24], while the situation with $\alpha_X$ is more complex. Indeed, as it follows from Eq.(6), that sign of $\alpha_X$ strongly depends on $U^*_m$ sign and $T_C$ value. In the case $\alpha_X<0$ Eq.(12b) could be used as trial function only when its denominator is positive finite quantity. Thus the critical thickness $l^X_{cr}(T,U_m)$ exists. It is the minimum root of equation $tg\left(l^X_{cr}/2l_X(T,U_m)\right)=l_X(T,U_m)/\lambda_X$.

Hereinafter we supposed that $E_{X,Y}=0$, $E_Z=E_0$. Integration in the expression (4) with the trial functions (11)-(12) leads to the following form of the free energy with renormalized coefficients:

$$\Delta G(P_V)=\begin{bmatrix} \dfrac{A_Z}{2}P_{VZ}^2+\dfrac{B_Z}{4}P_{VZ}^4+\dfrac{C_Z}{6}P_{VZ}^6-P_{VZ}\left(E_0-E_m\right)+\dfrac{D_m}{3}P_{VZ}^3+\dfrac{H_m}{5}P_{VZ}^5- \\ -P_{VZ}\left(K_m\left(P_{VX}^2+P_{VY}^2\right)+L_mP_{VX}^2P_{VY}^2+F_m\left(P_{VX}^4+P_{VY}^4\right)\right)+\dfrac{A_X}{2}\left(P_{VX}^2+P_{VY}^2\right)+ \\ +\dfrac{B_X}{4}\left(P_{VX}^4+P_{VY}^4\right)+\dfrac{C_X}{6}\left(P_{VX}^6+P_{VY}^6\right)+\dfrac{C_{XY}}{2}P_{VX}^2P_{VY}^2+\dfrac{F_{XZ}}{2}P_{VZ}^2\left(P_{VX}^2+P_{VY}^2\right)+ \\ +\dfrac{C_{XYZ}}{2}P_{VX}^2P_{VY}^2P_{VZ}^2+\dfrac{F_{XXX}}{4}\left(P_{VX}^4P_{VY}^2+P_{VY}^4P_{VX}^2\right)+\dfrac{F_{XYZ}}{4}P_{VZ}^4\left(P_{VX}^2+P_{VY}^2\right)+ \\ +\dfrac{Q_{XYZ}}{4}P_{VZ}^2\left(P_{VX}^4+P_{VY}^4\right) \end{bmatrix} \tag{13}$$

The approximate expressions for the renormalized coefficients in Eq.(13) for the second order phase transitions are derived in Appendix A. Because in the majority of cases misfit strain is very small $U_m(T) << 0.1$, only the linear on $U_m$ terms must be taken into account in Eq.(13). Neglecting the terms proportional to $P_m^2/P_s^2$ and this ratio higher powers, and introducing the dimensionless parameters $h = l/2\,l_Z$, $h_d = l_d/2\,l_Z$, $\Lambda_Z = \lambda_Z/\,l_Z$ the renormalized coefficients acquire the following form:

$$A_Z = \frac{\alpha_Z + 4\pi\overline{\varphi(z)}}{1-\overline{\varphi(z)}} \approx \alpha_T\left(T - T_C - \frac{2Q_{12}U_m^*}{\alpha_T(S_{11}+S_{12})} + \frac{4\pi}{\alpha_T}\cdot\psi(h)\right) \qquad (14)$$

hereinafter

$$\overline{\varphi(z)} = \psi(h) = \frac{sh(h)}{h\left(ch(h) + \Lambda_Z\, sh(h)\right)} \approx \begin{cases} \dfrac{1}{h(1+\Lambda_Z)}, & h >> 1 \\ \dfrac{1}{1+\Lambda_Z h}, & h << 1 \end{cases} \qquad (15)$$

It is seen that always $0 < \psi(h) < 1$ at $\Lambda_Z > 0$, therefore introduced in denominator of Eq.(14) multiplier $(1-\psi(h))$ is positive and could not be zero, thus it does not change $A_Z$ sign. For the case $P_Z \neq 0,\ P_X = P_Y = 0$, the condition $\left(\alpha_Z + 4\pi\overline{\varphi(z)}\right) = 0$ gives the critical temperature $T_{cr}^Z(h, U_m)$ (at arbitrary thickness) or thickness $h_{cr}^Z(T, U_m)$ (at arbitrary temperature $T < T_C^Z$) where paraelectric phase loses its stability independently on phase transition order. $T_{cr}^Z(h, U_m)$ and $h_{cr}^Z(T, U_m)$ coincide with critical parameters for the second order phase transitions. The introduced critical parameters strongly simplify the expression for the coefficient $A_Z$, namely:

$$A_Z(T, h, U_m) \approx \alpha_T\left(T - T_{cr}^Z(h, U_m)\right), \qquad (16)$$

Expressions for $T_{cr}^Z(h, U_m)$ could be written as:

$$T_{cr}^Z(h, U_m) \approx T_C\left(1 + \frac{2Q_{12}U_m^*(h)}{\alpha_T T_C(S_{11}+S_{12})} - \frac{4\pi}{\alpha_T T_C}\cdot\psi(h)\right)$$
$$U_m^*(h) \approx \begin{cases} U_m, & h \leq h_d \\ U_m\dfrac{h_d}{h}, & h > h_d \end{cases} \qquad (17)$$

Note, that for $h \leq h_d$ and $h(1+\Lambda_Z) >> 1$ Eq.(16) coincides with expression for $T_{cr}^Z$ obtained in [14]. At $\psi(h) << 1$ (i.e. $h(1+\Lambda_Z) >> 1$) one can obtain from Eqs (14), (15) that at $T < T_C^Z$:

$$A_Z(T, h, U_m) \approx \frac{4\pi}{1+\Lambda_Z}\left(\frac{1}{h} - \frac{1}{h_{cr}^Z(T, U_m)}\right). \qquad (18)$$

$$h_{cr}^Z(T, U_m) = \begin{cases} \dfrac{4\pi}{\alpha_T\left(T_C + 2Q_{12}U_m/\alpha_T(S_{11}+S_{12}) - T\right)\,(1+\Lambda_Z)}, & h_{cr}^Z \leq h_d(U_m) \\ \dfrac{4\pi}{\alpha_T(T_C - T)\,(1+\Lambda_Z)} - \dfrac{2Q_{12}U_m h_d}{\alpha_T(T_C - T)(S_{11}+S_{12})}, & h_{cr}^Z > h_d(U_m) \end{cases} \qquad (19)$$

It is seen from Eq.(19) that $h_{cr}^Z(0, U_m < 0) < h_{cr}^Z(0, U_m = 0) < h_{cr}^Z(0, U_m > 0)$ allowing for $Q_{12} < 0$ in perovskites. Therefore the critical thickness is maximal for tensile strain $U_m > 0$ and minimal for compressive strain $U_m < 0$. Mismatch-induced internal field:

$$E_m = 2\pi\overline{\varphi(z)}\cdot P_m \approx \frac{2\pi U_m}{d_{31}}\psi(h). \qquad (20)$$

All other coefficients before $P_Z$ powers have rather simple form only at $h(1+\Lambda_Z) >> 1$:

$$B_Z \approx \beta_Z, \qquad C_Z \approx \gamma_Z, \tag{21}$$

$$D_m \approx -\frac{3\beta_Z U_m}{2d_{31}}\cdot\frac{1}{h(1+\Lambda_Z)}, \quad H_m \approx -\frac{5\gamma_Z U_m}{2d_{31}}\cdot\frac{1}{h(1+\Lambda_Z)}, \tag{22}$$

The coefficients before $P_{X,Y}$ powers obtained with the help of trial function (11)-(12) have rather simple form only at $(\lambda_X/l_X)^2 >> 1$ or $(\lambda_X/l_X)^2 << 1$ (see (A.8)), namely:

$$A_X(T,h,U_m) = \frac{\alpha_X}{1-\phi(z)} \approx \alpha_T\left(T - T_{cr}^X(h,U_m)\right), \tag{23}$$

Expressions for $T_{cr}^X(h,U_m)$ could be written as:

$$T_{cr}^X(h,U_m) \approx \begin{cases} T_C^X(U_m^*) - \dfrac{\pi^3(\delta_X/\delta_Z)}{\alpha_T}\cdot\dfrac{1}{h^2}, & \left(\dfrac{\lambda_X}{l_X}\right)^2 << 1 \\ T_C^X(U_m^*) - \dfrac{\delta_X}{\alpha_T\lambda_X l_Z}\cdot\dfrac{1}{h}, & \left(\dfrac{\lambda_X}{l_X}\right)^2 >> 1 \end{cases} \tag{24 a,b}$$

The definition of $T_C^X(U_m^*)$ is given in Eq.(6). At $T < T_C^X(U_m^*)$ the critical thickness $h_{cr}^X(T,U_m)$ can be introduced in the expression for the coefficient $A_X$, namely:

$$A_X(T,h,U_m) \approx \begin{cases} \dfrac{\pi^3\delta_X}{\delta_Z}\left(\dfrac{1}{h^2} - \dfrac{1}{h_{cr}^X(T,U_m)^2}\right), & h_{cr}^X \approx \sqrt{\dfrac{\pi^3\delta_X/\delta_Z}{\alpha_T\left(T_C^X(U_m^*)-T\right)}}, \quad \left(\dfrac{\lambda_X}{l_X}\right)^2 << 1 \\ \dfrac{\delta_X}{\lambda_X l_Z}\left(\dfrac{1}{h} - \dfrac{1}{h_{cr}^X(T,U_m)}\right), & h_{cr}^X \approx -\dfrac{\left(\delta_X/\lambda_X l_Z\right)}{\alpha_T\left(T - T_C^X(U_m^*)\right)}, \quad \left(\dfrac{\lambda_X}{l_X}\right)^2 >> 1 \end{cases} \tag{25a,b}$$

Note, that for the majority of cases $\left(\lambda_X/l_X\right)^2 = |\alpha_X|\lambda_X^2/\delta_X \le 10^{-2}\cdot 10^2 nm^2/10^{-15}cm^2 << 1$, so that in what follows we will use (24a) and (25a). For the case $P_Z = 0$, $P_{X,Y} \ne 0$, the condition $A_X = 0$ gives the critical temperature $T_{cr}^X(h,U_m)$ (at arbitrary thickness) or thickness $h_{cr}^X(T,U_m)$ (at arbitrary temperature $T < T_C^X$) where paraelectric phase for transverse components $P_{X,Y}(z)$ loses its stability independently on phase transition order.

Other coefficients at $h(1+\Lambda_Z) >> 1$ have the form (see Appendix A):

$$K_m \approx \frac{\eta_Z}{2}\frac{U_m}{d_{13}}\frac{1}{h(1+\Lambda_Z)}, \qquad L_m \approx \frac{\gamma_{XYZ}}{2}\frac{U_m}{d_{13}}\frac{1}{h(1+\Lambda_Z)}, \qquad F_m \approx \frac{\eta_{XYZ}}{2}\frac{U_m}{d_{13}}\frac{1}{h(1+\Lambda_Z)} \tag{26}$$

$$B_X \approx \beta_X,\ C_{XY} \approx \eta_X,\ C_X \approx \gamma_X,\ F_{XZ} \approx \eta_Z,\ C_{XYZ} \approx \gamma_{XYZ},\ F_{XXX} \approx F_{XYZ} \approx Q_{XXX} \approx \eta_{XYZ}. \tag{27}$$

Note, that odd power $P_{VZ}^3$, $P_{VZ}\left(P_{VX}^2 + P_{VY}^2\right)$ in Eq.(13) is unusual for cubic symmetry perovskite structure ferroelectrics, these terms as well as $E_m$ are absent at $P_m = 0$ (see Eqs.(16), (17), (21)), i.e. they are related to mismatch effect. The obtained analytical expressions for renormalized coefficients will simplify the calculations of phase diagrams and properties of the films. In the following consideration we will pay special attention to the influence of internal misfit-induced field $E_m \sim U_m$ on the film properties and phase diagram.

## 4. FILMS PHASE DIAGRAMS AND DIELECTRIC PROPERTIES

### 4.1. General equations for phase diagrams calculations

The coupled equations for the amplitudes $P_V$ can be obtained by variation of the renormalized free energy (13). For the second order phase transitions this yields the following coupled equations:

$$\begin{cases}\left[A_Z + F_{XZ}\left(P_{VX}^2 + P_{VY}^2\right)\right]P_{VZ} + B_Z P_{VZ}^3 + D_m P_{VZ}^2 - K_m\left(P_{VX}^2 + P_{VY}^2\right) = E_0 - E_m,\\ \left[A_X + C_{XY}P_{VY}^2 + F_{XZ}P_{VZ}^2 - 2K_m P_{VZ}\right]P_{VX} + B_X P_{VX}^3 = 0,\\ \left[A_X + C_{XY}P_{VX}^2 + F_{XZ}P_{VZ}^2 - 2K_m P_{VZ}\right]P_{VY} + B_X P_{VY}^3 = 0.\end{cases} \tag{28}$$

The last two equations for $P_{VX,Y}$ can be solved independently and their solutions $P_{VX,Y}(P_{VZ})$ must be substituted into the first equation for $P_{VZ}$ as well as into the Eq.(13). Similarly to the bulk ferroelectrics, these equations allowed us to analyze possible equilibrium orientations of polarization for some values of the coefficients and so for some film thickness, temperature and misfit strain regions. By this way we obtained different phases with free energies $G_{A,S,C}(P_{VZ})$, namely:

a). "Asymmetrical" *ac*-phase ($P_X \neq 0 \leftrightarrow P_Y = 0,\quad P_Z \neq 0$) and *a*-phase ($P_X \neq 0,\quad P_{Y,Z} = 0$):

$$P_{VX} \equiv 0,\quad P_{VY}^2 = -\frac{A_X + F_{XZ}P_{VZ}^2 - 2K_m P_{VZ}}{B_X},\quad G_A(P_{VZ}) = \min, \tag{29a}$$

$$G_A(P_{VZ}) = \left[\begin{array}{l}\left(A_Z - \dfrac{F_{XZ}A_X - 2K_m^2}{B_X}\right)\dfrac{P_{VZ}^2}{2} + \left(D_m + \dfrac{3K_m F_{XZ}}{B_X}\right)\dfrac{P_{VZ}^3}{3} + \\ + \left(B_Z - \dfrac{F_{XZ}^2}{B_X}\right)\dfrac{P_{VZ}^4}{4} - P_{VZ}\left(E_0 - E_m\right) - \dfrac{A_X^2}{4B_X}\end{array}\right]. \tag{29b}$$

b). Paraelectric ***PE*-phase** ($P_X = P_Y = P_Z = 0$):

$$\min\left[G_A(P_{VZ})\right] \geq 0 \;\&\; \min\left[G_S(P_{VZ})\right] \geq 0 \;\&\; \min\left[G_C(P_{VZ})\right] \geq 0. \tag{30}$$

*PE*-phase region corresponds to the conditions: $A_Z \geq 0$ & $A_X \geq 0$.

c). "Unipolar" ferroelectric ***FE<sub></sub>*** 

$$P_{VX} = P_{VY} = 0,\quad P_{VZ} \neq 0,\quad G_C(P_{VZ}) = \min, \tag{31a}$$

$$G_C(P_{VZ}) = \left[A_Z\frac{P_{VZ}^2}{2} + D_m\frac{P_{VZ}^3}{3} + B_Z\frac{P_{VZ}^4}{4} - P_{VZ}\left(E_0 - E_m\right)\right]. \tag{31b}$$

The condition of $FE_c$-phase absolute stability $\left\{\min\left[G_C(P_{VZ})\right] < 0 \;\&\; \min\left[G_C(P_{VZ})\right] < \min\left[G_{S,A}(P_{VZ})\right]\right\}$ determines the region of parameters $(U_m, T, h)$ where the unipolar phase of the film $\{P_{X,Y} \equiv 0,\quad P_Z \neq 0\}$ is energetically preferable. Boundary between $FE_c$-phase and *PE*-phase is $T = T_{cr}^Z(h, U_m)$ & $T \geq T_{cr}^X(h, U_m)$.

d). "Symmetrical" ferroelectric **$FE_r$-phase** ($P_X = P_Y \neq 0,\quad P_Z \neq 0$) and **$FE_{aa}$- phase** ($P_X = P_Y \neq 0,\quad P_Z = 0$):

$$P_{VX}^2 = P_{VY}^2 = -\frac{A_X + F_{XZ}P_{VZ}^2 - 2K_m P_{VZ}}{B_X + C_{XY}},\quad G_S(P_{VZ}) = \min, \tag{32a}$$

$$G_S(P_{VZ}) = \left[\begin{array}{l}\left(A_Z - \dfrac{F_{XZ}A_X - 2K_m^2}{B_X + C_{XY}}\right)\dfrac{P_{VZ}^2}{2} + \left(D_m + \dfrac{3K_m F_{XZ}}{B_X + C_{XY}}\right)\dfrac{P_{VZ}^3}{3} + \\ + \left(B_Z - \dfrac{F_{XZ}^2}{B_X + C_{XY}}\right)\dfrac{P_{VZ}^4}{4} - P_{VZ}\left(E_0 - E_m\right) - \dfrac{A_X^2}{2\left(B_X + C_{XY}\right)}\end{array}\right]. \tag{32b}$$

Boundary between *PE* and $FE_{aa}$-phase is $T = T_{cr}^{X}(h,U_m)$ & $T \geq T_{cr}^{Z}(h,U_m)$. Regions of $FE_r$ and $FE_c$-phase correspond to the conditions: $A_Z - \frac{F_{XZ}A_X - 2K_m^2}{B_X + C_{XY}} \leq 0$ and $A_Z \leq 0$ . The boundary of $FE_r$-phase $T_{cr}^{R}(h,U_m)$ can be obtained from the equation $A_Z - \frac{F_{XZ}A_X - 2K_m^2}{B_X + C_{XY}} = 0$, namely

$$\left(T_{cr}^{R} - T_{cr}^{Z}(h,U_m)\right) - \frac{F_{XZ}}{B_X + C_{XY}}\left(T_{cr}^{R} - T_{cr}^{X}(h,U_m)\right) = 0. \tag{33}$$

The critical thickness $h_{cr}^{R}(T,U_m)$ can be found from this equation by substitution $T$ for $T_{cr}^{R}$ and $h_{cr}^{R}$ for $h$.

Under the temperature decrease *PE*-phase transforms into the one of ferroelectric phases with the highest transition temperature. Therefore this transition takes place at critical temperature $T_{cr}(h,U_m) = \max\left\{T_{cr}^{Z}(h,U_m), T_{cr}^{R}(h,U_m), T_{cr}^{X}(h,U_m)\right\}$. Since this transition temperature decreases with thickness decrease it is seen that the critical thickness can be found from condition $h_{cr}(T,U_m) = \min\left\{h_{cr}^{Z}(T,U_m), h_{cr}^{R}(T,U_m), h_{cr}^{X}(T,U_m)\right\}$.

## 4.2. Dielectric permittivity and pyroelectric coefficient

In accordance with definition [24], the average static linear dielectric permittivity $\overline{\varepsilon}_{zz} = 1 + 4\pi\, d\overline{P}_{VZ}/dE_0\big|_{E_0=0}$ can be obtained from Eq.(28). For example in ***FE<sub></sub>*** phase $\overline{\varepsilon}_{zz}$ could be found from the system of equations:

$$\begin{cases} \overline{\varepsilon}_{zz}(T,h) = \dfrac{4\pi}{A_Z + 3B_Z P_{VZ}^2 + 2D_m P_{VZ}} \\ A_Z P_{VZ} + D_m P_{VZ}^2 + B_Z P_{VZ}^3 + E_m = 0 \end{cases}. \tag{34}$$

Finally let us consider the pyroelectric coefficient z-component $\Pi_Z(z,T) = -\left(\frac{1-\varphi(z)}{1-\overline{\varphi(z)}}\right)\cdot\frac{dP_{VZ}(z,T)}{dT}$, which determines the measurable pyroelectric current $j_z$. Coefficient $\overline{\Pi_Z(h,T)}$ exists when $P_{VZ} \neq 0$. In our approximations functions $\varphi(z)$, $\xi(z)$ and coefficients $B_{Z,X}$, $D_m$, $E_m$ are almost independent on temperature and so only $dA_Z(T)/dT = \alpha_T$ and $dA_X(T)/dT = \alpha_T$ will contribute to $d\overline{P}_Z(z,T)/dT$. For example in $FE_c$ phase $\overline{\Pi_Z(h,T)}$ could be found from the system of equations:

$$\begin{cases} \overline{\Pi_Z(h,T)} \approx \alpha_T \dfrac{P_{VZ}}{A_Z + 2D_m P_{VZ} + 3B_Z P_{VZ}^2} \\ A_Z P_{VZ} + D_m P_{VZ}^2 + B_Z P_{VZ}^3 = E_0 - E_m \end{cases} \tag{35}$$

## 4.3 The properties and phase diagrams for $PbZr_{0.5}Ti_{0.5}O_3$ (PZT50/50) films on different substrates

The temperature dependences of linear dielectric permittivity $\overline{\varepsilon}_{zz}$ calculated from Eq.(34) at $E_0 = 0$ for different film thickness values $h = l/2l_Z$ are depicted in Figs 1a-4a. It is seen that linear permittivity diverges ($U_m = 0$) or it has maximum ($U_m \neq 0$) at $T = T_{cr}(h,U_m)$. The maximum diffuse with film thickness decrease and then disappear at $h < h_{cr}(T = 0)$. It is worth to draw attention to the increase of permittivity value with temperature decrease at $h < h_{cr}(T = 0)$ (see curves for the lowest thickness values in Figs 1a-4a).

The temperature dependences of pyroelectric coefficient z-component $\overline{\Pi_Z(h,T)}$ calculated from Eq.(35) at $E_0 = 0$ for different film thickness values are depicted in Figs 1b-4b. It is seen, that only if $U_m = 0$, $\overline{\Pi_Z(h,T)}$ diverges at $T = T_{cr}(h,U_m)$ and then completely disappears in *PE*-phase at $T > T_{cr}(h,U_m)$ (see Fig.3b). In all other cases ($U_m \neq 0$) pyroelectric coefficient $\overline{\Pi_Z(h,T)}$ exists at $T > T_{cr}(h,U_m)$, at $T = T_{cr}(h,U_m)$ it has only maximum, which diffuses at film thickness *h* decreasing and then disappears at $h < h_{cr}(T = 0)$. The position of $\overline{\Pi_Z(h,T)}$ maximum is close to the position of $\overline{\varepsilon}_{zz}$ maximum. At $h < h_{cr}(T = 0)$ the behavior of pyroelectric coefficient looks like that of $\overline{\varepsilon}_{zz}$, namely it increases with temperature decrease (see curves for the lowest thickness values in Figs 1b-4b).

So, the critical thickness can be determined from the measurements of the linear dielectric permittivity or pyroelectric coefficient temperature dependences. Namely, this is the thickness below which their temperature maxima disappear. Nonzero value of $\overline{\Pi_Z(h,T)}$ at $h < h_{cr}(T = 0)$ is related to the built-in electric field $E_m \sim U_m$.

However, despite strained films have misfit-induced polarization $P_{VZ}$ and $\Pi_Z$ practically in all thickness region, their hysteresis loops shape is significantly different at $h < h_{cr}$ and $h > h_{cr}$. Namely, only above the critical thickness hysteresis loops have some width $E_C = \left(E_C^+ - E_C^-\right)/2 \neq 0$ related to the left and right hand side values of coercive fields:

$$E_C(T,h) \approx \frac{2}{3}\sqrt{-\frac{\alpha_T^3\left(T - T_{cr}(h,U_m)\right)^3}{3B_Z}} + E_m. \qquad (36)$$

While below the critical thickness we obtain only asymmetrical curve without hysteresis loop shifted on $E_m$ value ($E_C^+ = E_C^- = E_m$) [14].

Aforementioned peculiarities of the properties behavior at $h < h_{cr}$ are related to the influence of built-in field $E_m$ and they resemble the behavior of electrets [20]. Thus, for the thinnest films with thickness below critical value, built-in field $E_m$ induces **electret-like polar state** that possesses non centrosymmetrical structure, reveal pyroelectric effect but has no ferroelectric hysteresis, domain structure and does not show a phase transition into nonpolar state with temperature increase.

In the Fig.1c-4c we give the results of the phase diagrams calculations. For the sake of comparison we considered two cases: $E_0 = E_m$ (internal field is compensated by external field) and $E_0 = 0$ (internal field influence the properties). First of all we would like to underline, that at all values $U_m \neq 0$ *PE* phase and so $h_{cr}$ (as the thickness that corresponds to the transition from *FE* to *PE*-phase) exist only at $E_0 = E_m$. At $E_0 \neq E_m$ or $E_0 = 0$ and $U_m \neq 0$ uncompensated difference $\left(E_0 - E_m\right)$ induces the polarization $P_{VZ}$ in thin films, so that *PE*-phase disappears and electret state arises at $h < h_{cr}$ while $FE_{aa}$-phase transforms into $FE_r$-phase (see Fig.1c). So that for the real thin film-substrate pair with $U_m \neq 0$ there is no transition from ferroelectric to the paraelectric phase, but to the electret state at critical thickness $h = h_{cr}$.

Also one can see the slight change of the phase boundaries at $h > h_d$, when misfit dislocations appear and decrease the misfit strain.

Let us consider in more detail the influence of misfit strain $U_m$ on phase diagrams. The critical thickness $h_{cr}$ decreases at negative $U_m$ values in comparison with a freestanding film and the films with $U_m > 0$ (compare the Figs. 1c-4c).

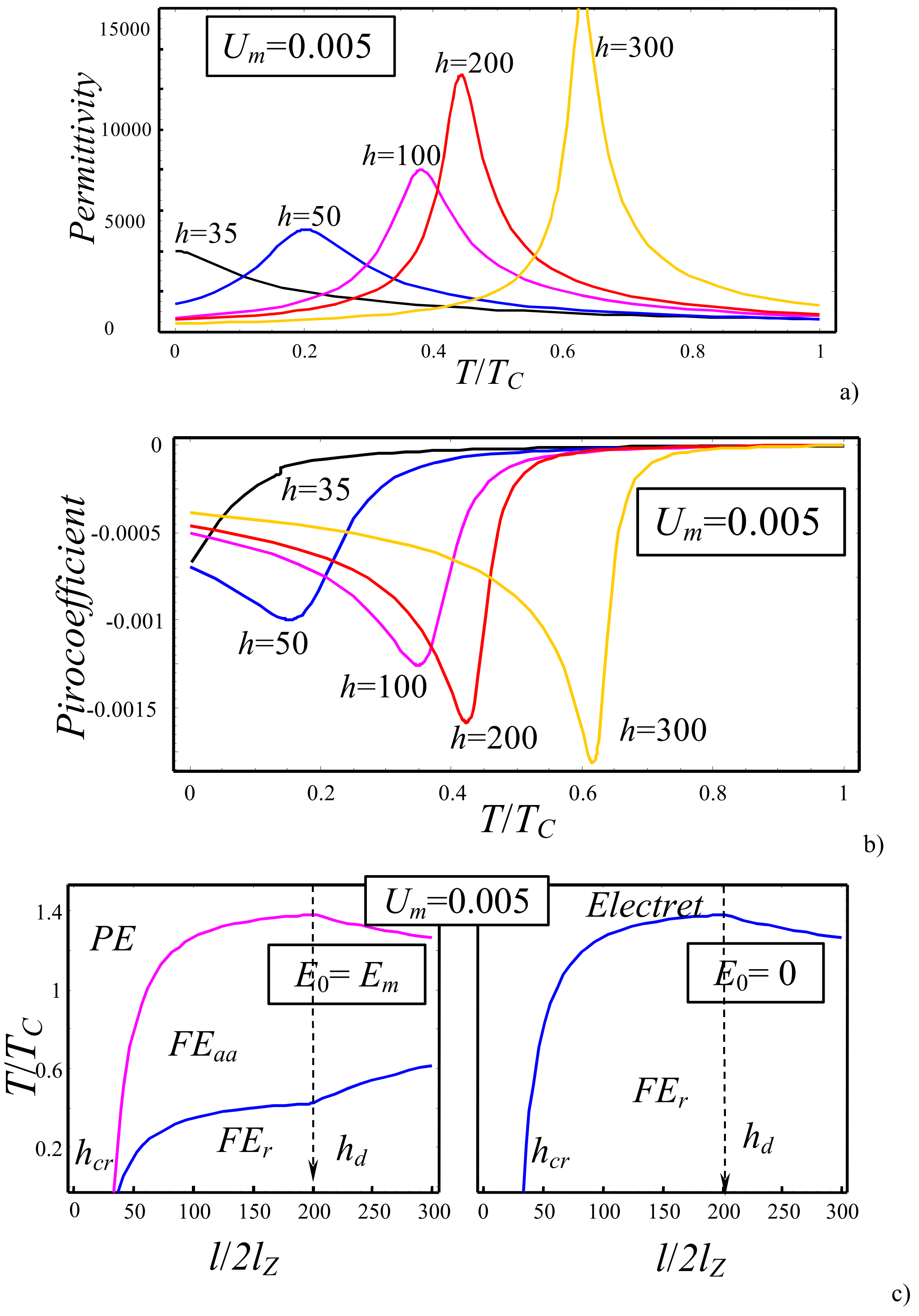

**Figure 1.** The dielectric permittivity $\varepsilon_{zz}(T)$, pyroelectric coefficient $\Pi(T)$ and phase diagram $T(h)$ for the following parameters: $\Lambda_Z = 50$, $\delta_Z \approx \delta_X$, $h_d \approx 1/|U_m|$, $\alpha_X \lambda_X^2/\delta_X << 1$ and different $h = l/2l_Z$ values. For 50/50 PZT film $2l_Z \sim 0.5nm$ thus $\lambda_Z \sim 12.5nm$, $T_C = 666K$.

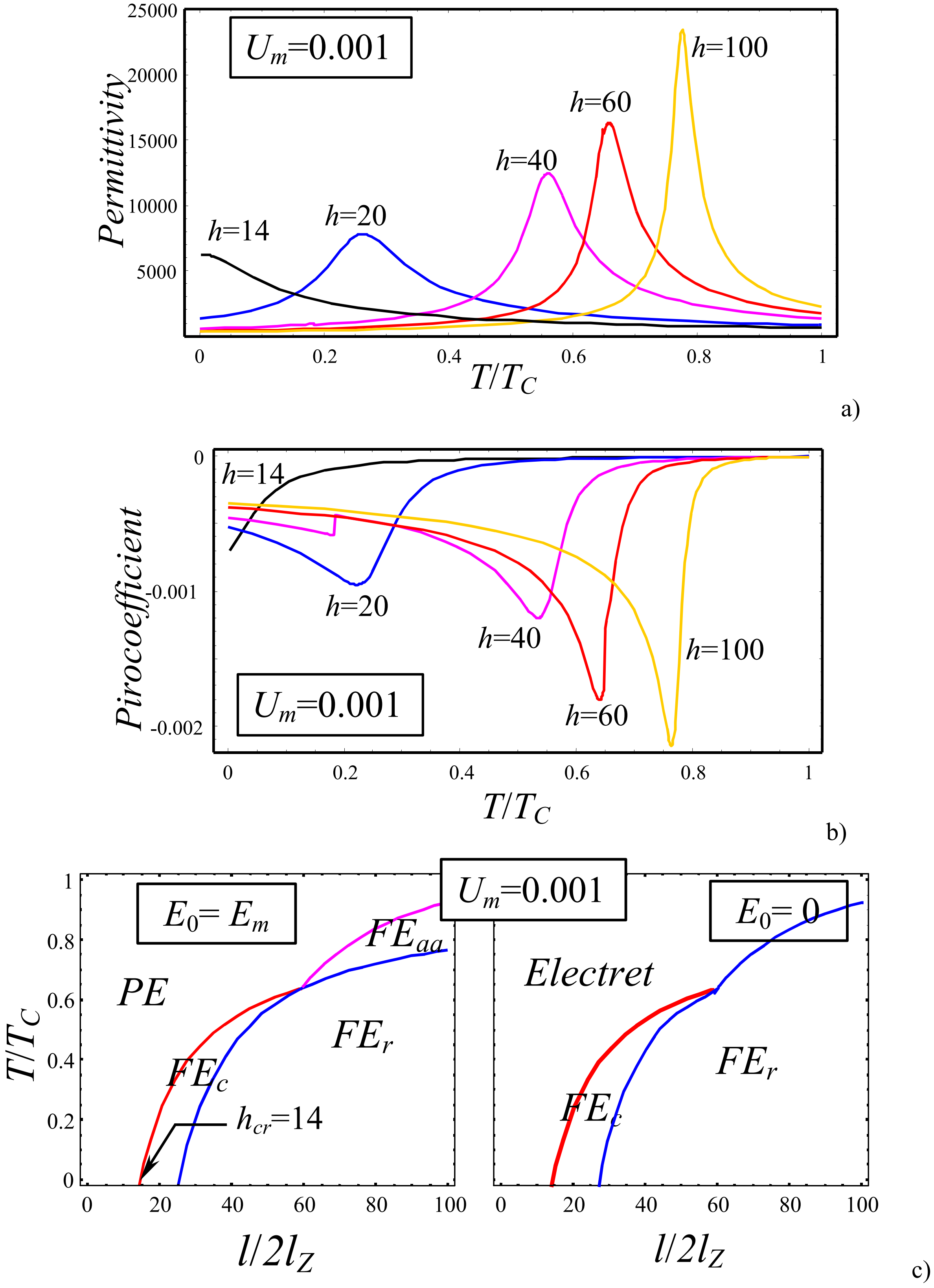

**Figure 2.** The phase diagram *T*(*h*), $\varepsilon_{zz}(T)$, $\Pi(T)$ for the following parameters: $\Lambda_Z = 50$, $\delta_Z \approx \delta_X$, $h_d \approx 1/|U_m|$, $\alpha_X \lambda_X^2/\delta_X << 1$ and different $h = l/2l_Z$ values. For 50/50 PZT film $2l_Z \sim 0.5nm$ thus $\lambda_Z \sim 12.5nm$, $T_C = 666\,K$.

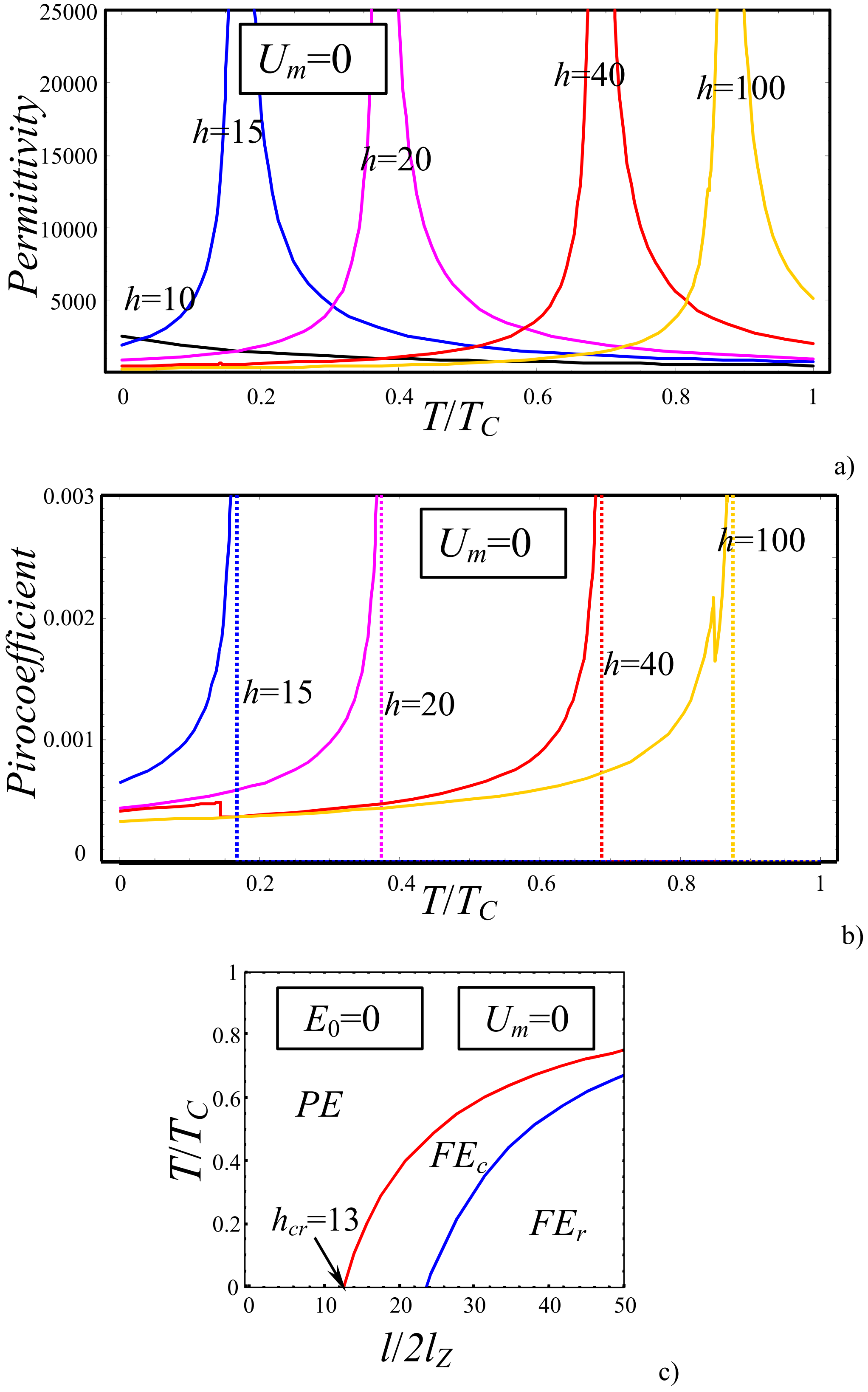

**Figure 3.** The phase diagram $T(h)$, $\varepsilon_{zz}(T)$, $\Pi(T)$ for the following parameters: $\Lambda_Z = 50$, $\delta_Z \approx \delta_X$, $h_d \approx 1/|U_m|$, $\alpha_X \lambda_X^2/\delta_X << 1$ and different $h = l/2l_Z$ values. For 50/50 PZT film $2l_Z \sim 0.5nm$ thus $\lambda_Z \sim 12.5nm$, $T_C = 666\,K$.

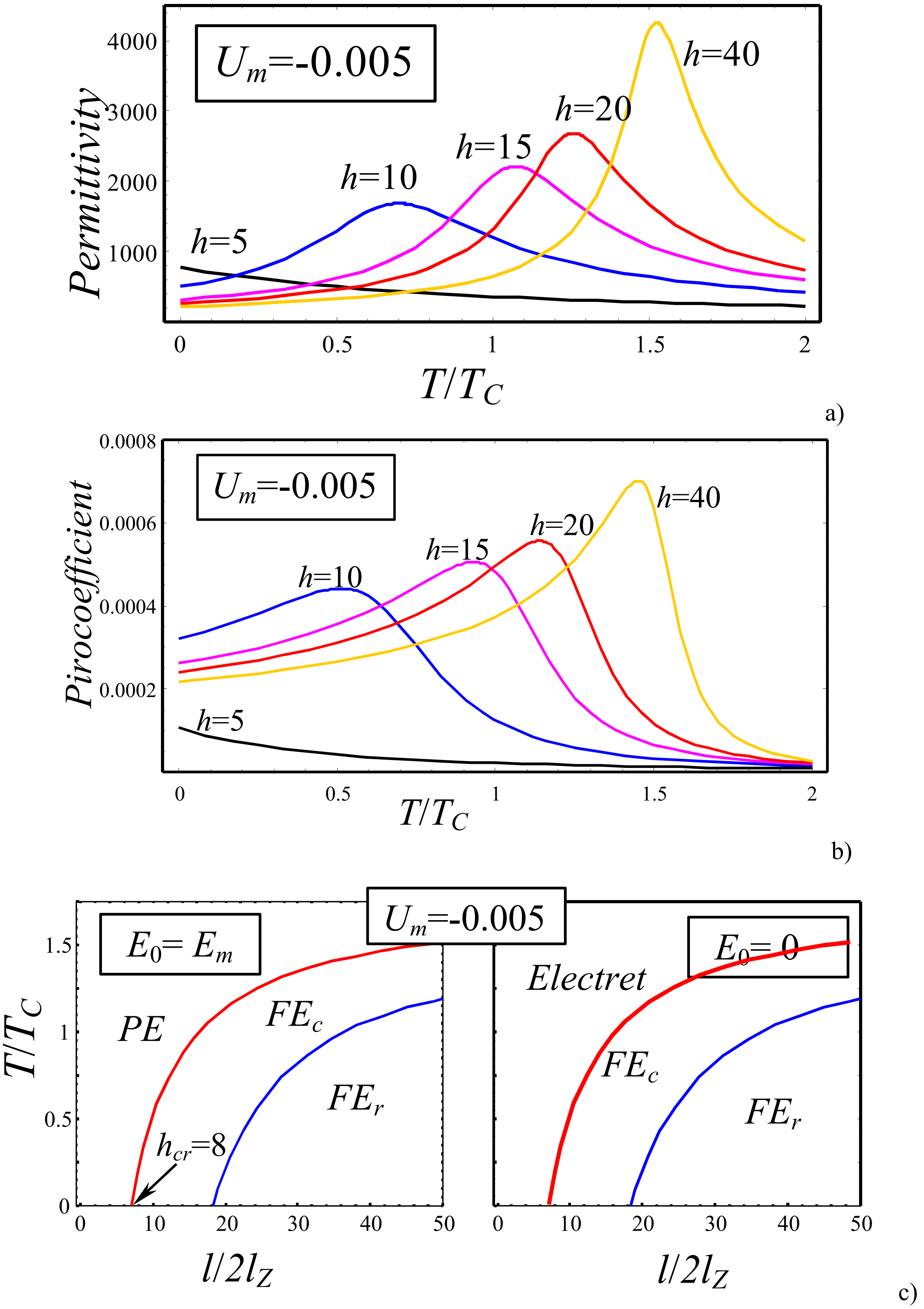

**Figure 4.** The phase diagram $T(h)$, $\varepsilon_{zz}(T)$, $\Pi(T)$ for the following parameters: $\Lambda_Z = 50$, $\delta_Z \approx \delta_X$, $h_d \approx 1/|U_m|$, $\alpha_X \lambda_X^2/\delta_X << 1$ and different $h = l/2l_Z$ values. For 50/50 PZT film $2l_Z \sim 0.5nm$ thus $\lambda_Z \sim 12.5nm$, $T_C = 666\,K$.

For $U_m = 0$ $FE_c$ phase exists along with $FE_c$ phase (see Fig. 3c). At $U_m = 0.001$ the region of $FE_c$ phase existence decreases and $FE_{aa}$ phase appears (see Fig. 2c). $FE_c$-phase almost disappears for $U_m > 0.0035$ so that only $FE_{aa}$ phase exists at $U_m = 0.005$ (see Fig.1c). For $U_m < 0$ the value of critical thickness essentially decreases and so it is not clear if size driven phase transition can exists. (see compare Figs 3c and 4c). Really our calculations based on Eq.(19) proved, that at $U_m \le -0.05$ critical thickness $h_{cr}$ is less than 1, i.e. it could be less than one lattice constant, because $2l_Z \approx 0.5nm$ for PZT and so $l_{cr}^Z < 0.5nm$. Therefore for such $U_m < 0$ value there is no thickness induced phase transition. Comparing phase diagrams in Figs 1c-4c, one can see that $FE_c$-phase occupies noticeable region mainly at $U_m \le 0$.

In any case $FE_c$-phase could exist only at definite $h, U_m, T$ values, namely at $h > h_{cr}(T, U_m)$. Taking into account that $E_m \sim U_m$ and $E_m$ is the built-in electric field, it always makes polarization values $+P_{VZ}$ and $-P_{VZ}$ energetically non-equivalent, so the spontaneous poling is possible. This becomes the most probable when misfit-induced field $E_m(h) \sim 1/h$ exceeds thermodynamic coercive field $E_C(h)$ from Eq. (36) for thin enough film. It is clear from the Figs 5a,b that for PZT(50/50) this occurs at $h < 25$ and $U_m \approx -0.001$ (see Fig.5a) or at $h < 22$ and $U_m \approx -0.005$ (see Fig.5b). So this phenomenon could explain of thin film spontaneous poling, because $E_m(h)$ could lead to the self-polarization of the film, up to the presence of the significant amount (~90%) of polarization in the normal to the surface direction (see Fig.5b) without any poling treatment. This self-polarization phenomenon has been observed by many authors (see [15], [16], [17]) and is known to be useful for applications.

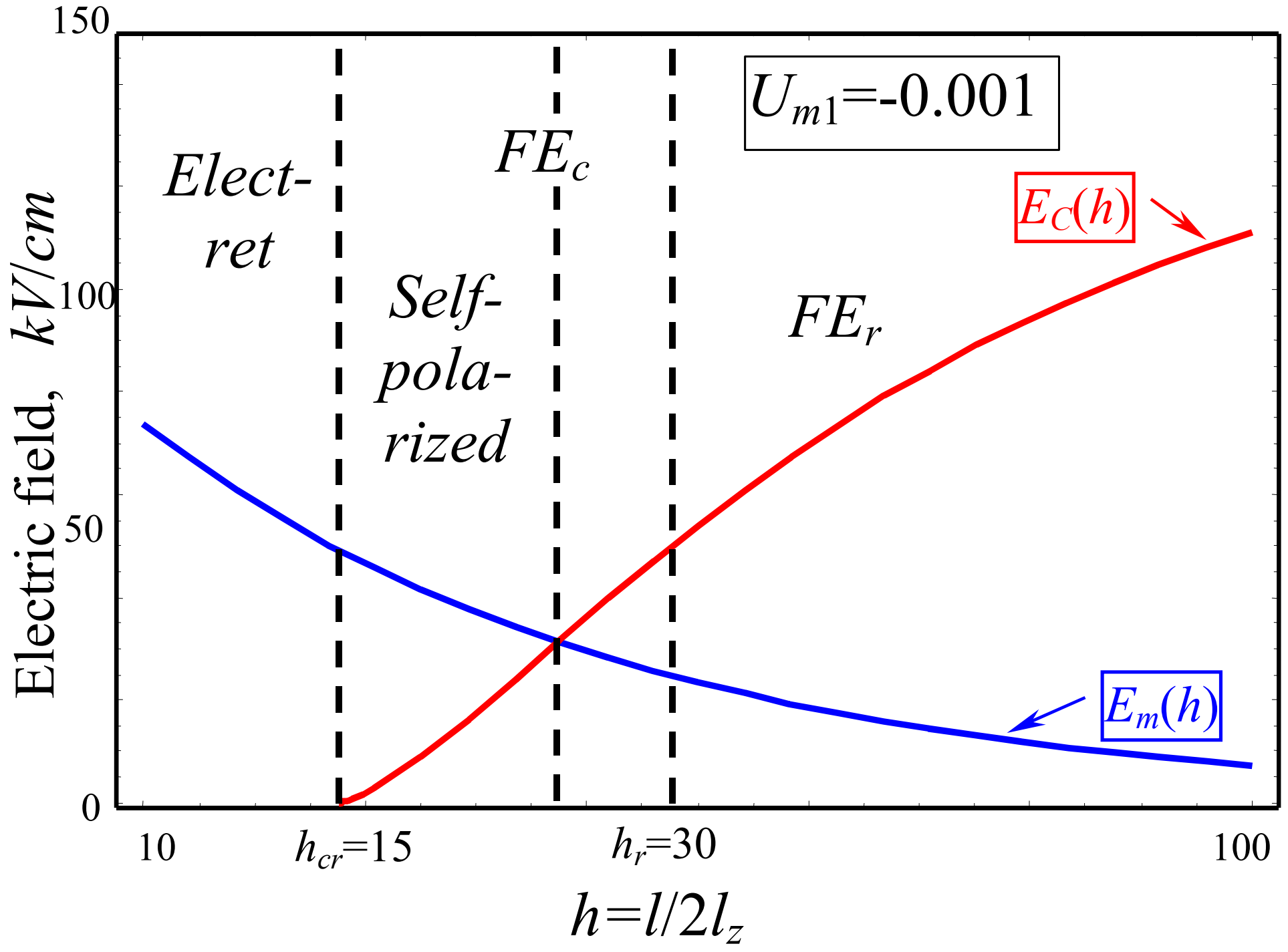

**Figure 5a.** Misfit-induced field $E_m(h)$ calculated for PZT(50/50) at $T = 300K$, $\Lambda_Z = 50$ in comparison with thermodynamic coercive field $E_C(h) \approx E_{0C}(1 - h_{cr}/h)^{3/2}$, for $U_{m1} = -0.001$ (a) amplitude $E_{0C}(U_{m1}) \approx 140\ kV/cm$ and for $U_{m1} = -0.005$ (b) $E_{0C}(U_{m2}) \approx 360\ kV/cm$.

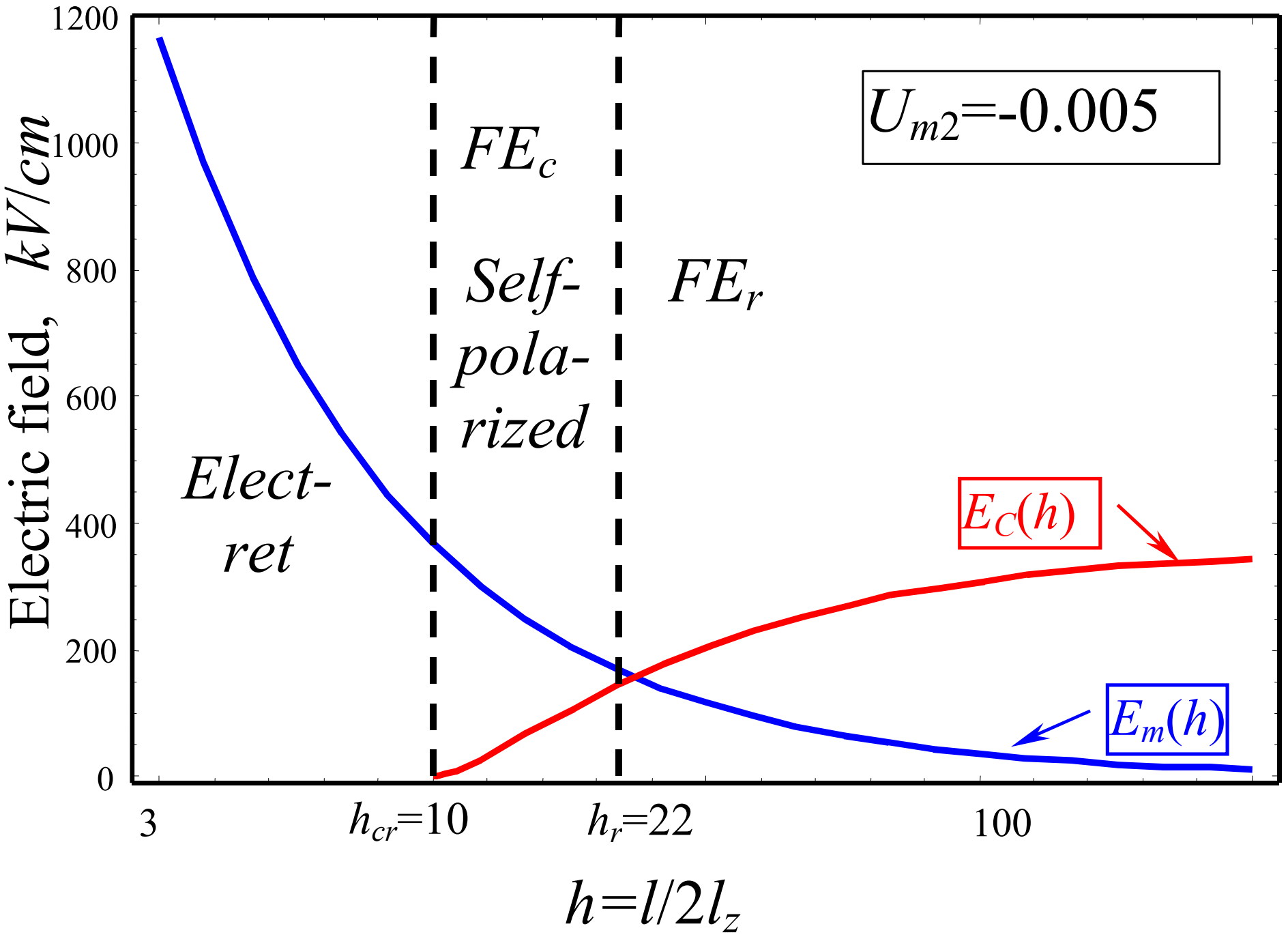

**Figure 5b**.

## 5. COMPARISON WITH EXPERIMENT AND DISCUSSION

Our theoretical calculations predict the conservation of spontaneous polarization at $h < h_{cr}$ due to the misfit-induced field $E_m$ in the strained ultrathin ferroelectric films. This means, that at $h < h_{cr}$ the tetragonality *c/a* depends not only on misfit between film and substrate as $\frac{c}{a} \approx 1 + \frac{s_{12} - s_{11}}{s_{11} + s_{12}} U_m^*$ [9], but also on the misfit-induced polarization $P_Z \sim E_m$ in accordance with Eqs.(13)-(28), namely:

$$\begin{cases} \frac{c}{a} \approx 1 + \frac{s_{12} - s_{11}}{s_{11} + s_{12}} U_m^* + \left( P_{VZ}(E_m) - \frac{U_m}{2d_{13}} \psi(h) \right)^2 \cdot \left( Q_{11} - \frac{2s_{12}}{s_{11} + s_{12}} Q_{12} \right), \\ A_Z(T, U_m, h) P_{VZ} + B_Z P_{VZ}^3 + C_Z P_{VZ}^5 + D_m P_{VZ}^2 + H_m P_{VZ}^4 = -E_m(U_m, h) \end{cases} \quad (36)$$

Let us compare (36) with the measured in [9] tetragonality of the system $PbTiO_3$ on $SrTiO_3$:Nb substrate. The authors of [9] reported about $a = 3.969A$, $b = 3.905A$ for pseudomorphic PTO on Nd-STO, so it posses misfit strain $U_m = -0.0164$ at room temperature. Our fitting for this filn *c/a*-ratio is presented in Fig.6. It is clear, that our approach that takes into account misfit-induced field (solid curve at $E_m \neq 0$), fits experimental data much better than the ones [12], [21] which do not consider surface piezoeffect (dashed curve at $E_m = 0$).

In accordance with our results neither paraelectric nor ferroelectric structure with polarization hysteresis could exist at $h < h_{cr}$, instead electret-like phase, where polarization has paraelectric-like dependence over internal field $E_m$ is expected.

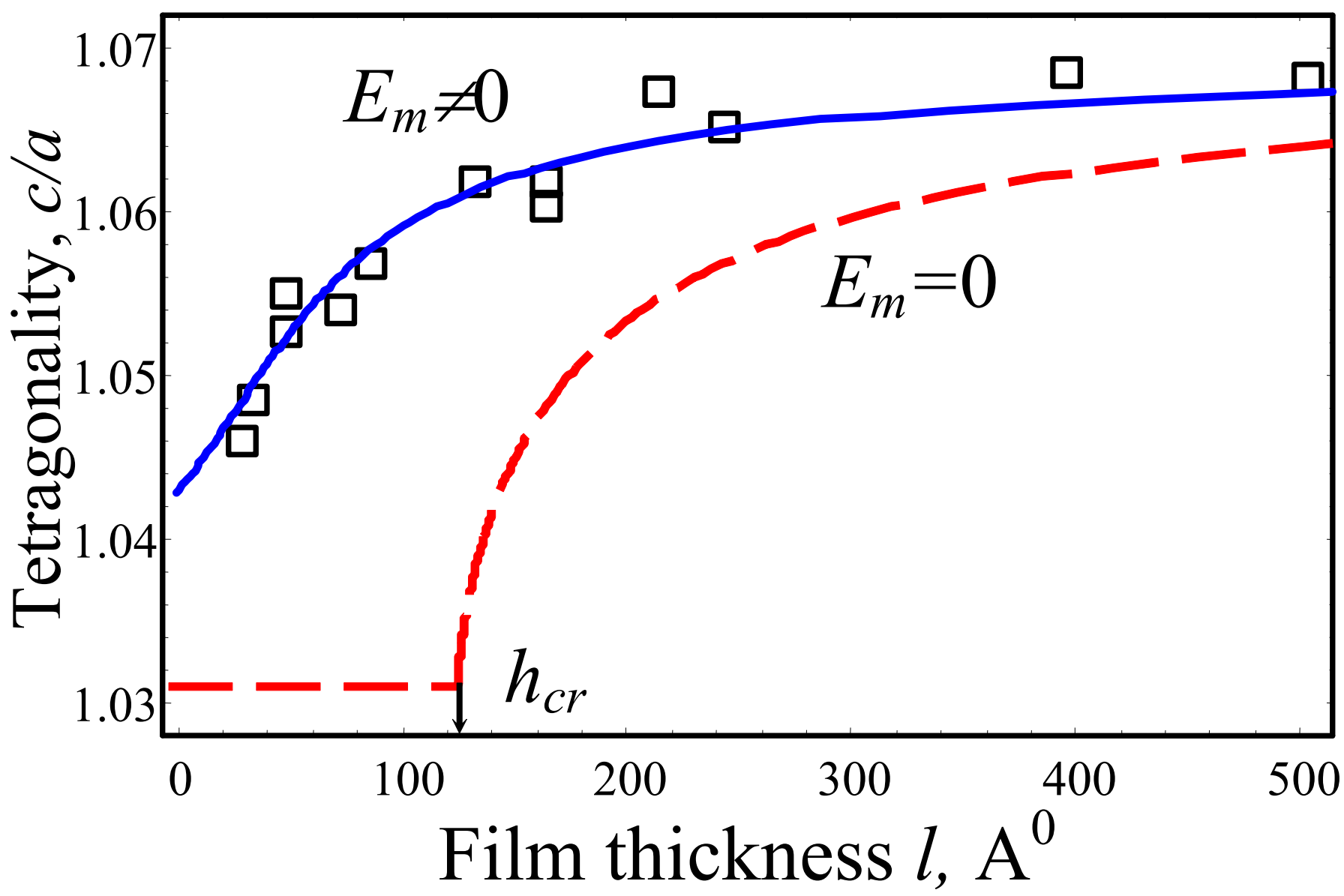

**Figure 6.** Tetragonality *c/a* via film thickness *l* for $PbTiO_3$ on $SrTiO_3$:Nb substrate. Squares are experimental data from [9], solid curve is our fitting at $T = 300K$, $\lambda_Z = 4.5nm$, $l_Z = 0.4nm$, $d_{31} = -19\cdot 10^{-3}\, Vm/N$, $U_m = -0.016$, dashed curve corresponds to the case $E_m = 0$ and the same other parameters.

The critical thickness $h_{cr}^Z$ decreases for negative $U_m$ values and rapidly increases for positive $U_m$ values in comparison with a free-standing film. Really, authors of [30] have obtained clear ferroelectric switching behavior for 300-200nm thick PZT films on GaN/sapphire substrate ($U_m > 0$), which is in contrast to their thinnest film (100nm), where no switching have been observed due to the probably built-in polarization. In contrast to this case the compressed thin films with $U_m < 0$ such as PZT(52/48) on Pt/$SrRuO_3$ substrates [31], [32], PZT(20/80) thin films on Nb-doped $SrTiO_3$ substrates [33] or PZT(35/65)/Ir [34] maintained their ferroelectric and piezoelectric properties at least up to 4-8nm thickness in the accordance with our predictions.

Recently authors of [8] have shown that ferroelectric domains exist in $PbTiO_3$ on $SrTiO_3$ substrate up to 3 unit cell thick film (i.e. x-ray domain scattering reveals fine structure of Bragg peaks that may be associated with c-domain structure). The system $PbTiO_3/SrTiO_3$ posses at temperatures $T \sim (750-850)K$ misfit strain $U_m \sim -0.01 \div -0.005$ (e.g. $a \approx 3.956A$, $b \approx 3.926A$ at 800K). One could easily estimate that for the ultrathin film with $l < l_d \le 20nm$ renormalized temperature $T_C^Z(U_m) = T_C + \dfrac{2Q_{12}U_m}{\alpha_T(S_{11}+S_{12})} \approx 930K$. The boundary between electret-like and $FE_c$ phases could be found from the Eq.(18a) as $T \approx T_C^Z\left(1-\dfrac{l_{cr}}{l}\right)$. Our fitting is represented in Fig.7 and one can see that approximately $l_{cr} \approx 0.75$ lattice constant, i.e. ferroelectricity could conserve even up to the monolayer film $PbTiO_3/SrTiO_3$.

Note, that to our mind the experimental data for the 3 unit cell film represents not $PbTiO_3$ on $SrTiO_3$ substrate, but rather $PbTiO_3$-$SrTiO_3$ solid solution 50/50. Since the bulk transition temperature for this materials is about 400K [27] one can see that for the strained film it rises approximately to 600K. Experimentally observed value of critical temperature 520K can be explained by size effects. It is worth to stress that the change of film composition due to the interpenetration of film and substrate or the appearance of the interface layer consisting film-substrate material solid solution can be expected for 1, 2 or 3 monolayers, so that one can hardly consider the results as the evidence in favor of ferroelectricity conservation in such ultra-thin film. One the other hand the phenomenological approach can be valid only for the thickness more than correlation radius, which is about a few lattice constants [22]. Because of this we did not try to fit the point for the film of 3 monolayers with our theory.

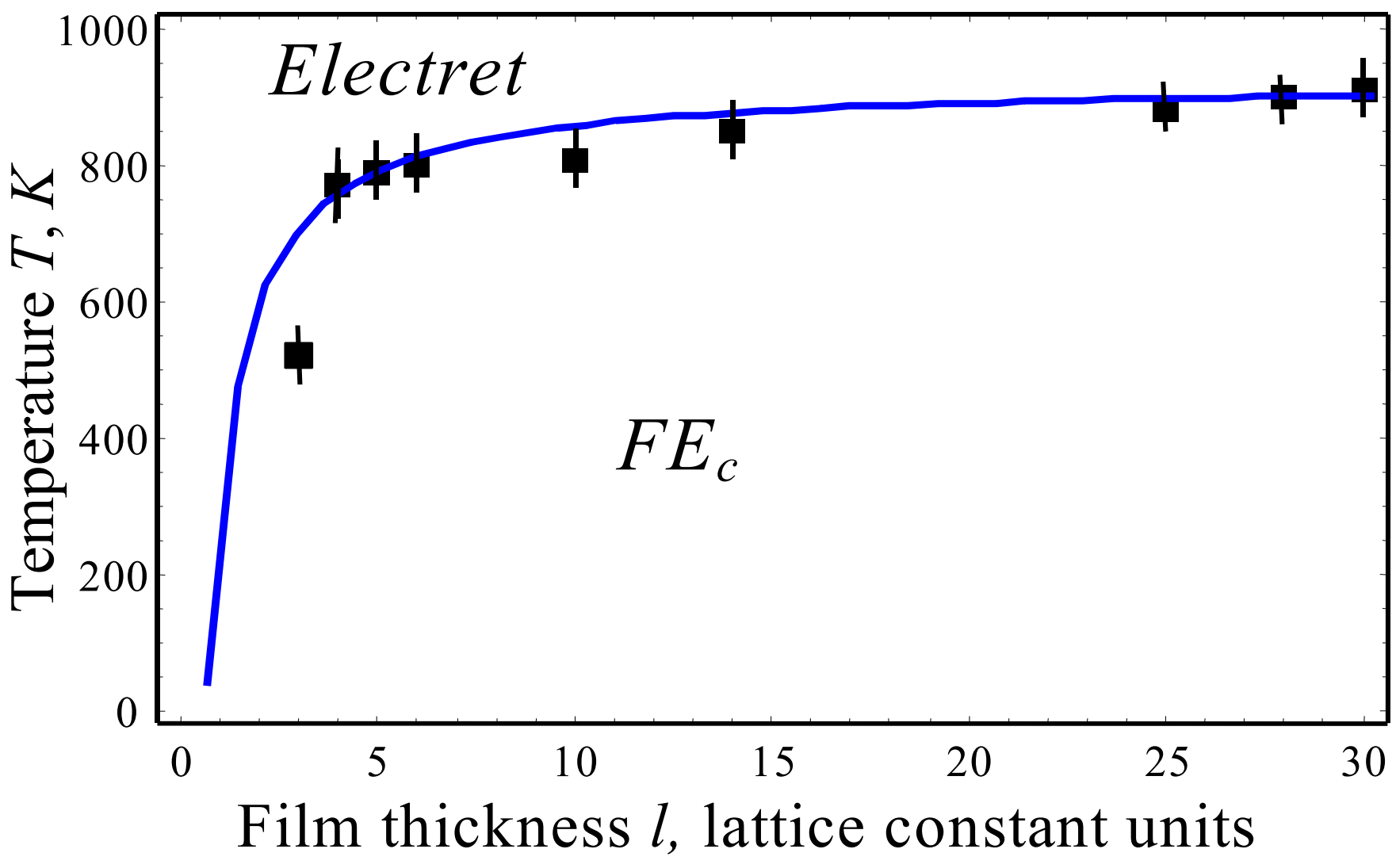

**Figure 7.** Phase diagram temperature versus film thickness for the films of $PbTiO_3$ on $SrTiO_3$ substrate. Squares are experimental data from [8], solid curve is our fitting with $U_m \approx -0.007$, $T_C^Z(U_m) = 930K$, $l_{cr}$=0.75 lattice constant.

## CONCLUSION

Three coupled Euler-Lagrange equations for $P_X$, $P_Y$, $P_Z$ were solved with the help of variational method, that allows obtaining free energy of polynomial type, but with renormalized coefficients. The analytical expression for the renormalized free energy coefficients of the perovskite ferroelectric films dependence on the misfit strain originated from the difference of the substrate and the film lattice constants and thermal expansion coefficients, depolarization field, correlation effects and surface piezoelectric effect caused by the absence of inversion centre in *z* direction near the surface even in the cubic phase were derived. This free energy allows to calculate the properties by conventional minimization procedure.

In the strained films the surface piezoeffect induces the internal electric field $E_m$, which shifts and diffuses the phase transition. For the thinnest films with thickness below critical value misfit-induced field $E_m$ induces electret like polar state with unswitchable polarization, its direction is determined by the misfit sign. Our theory predicts that mismatch-induced field $E_m$ could be comparable with thermodynamic coercive field and thus cause self-polarization in thin ferroelectric films.

With the help of this free energy we calculated the phase diagrams, i.e. we examine the influence of misfit strain on the critical temperatures and critical film thickness for the phase transitions from the electret state to the ferroelectric phase and between the ferroelectric phases with different symmetry allowed for the perovskite ferroelectrics ($FE_c$, $FE_r$) without external field. Namely at $E_0 = 0$:

- For the compressive strain ($U_m$<0) electret-like phase transforms into $FE_c$ –phase. The critical thickness of this transition $h_{cr}$ decreases with $|U_m|$ increase. Note, that the critical temperature firstly increases at $h<h_d$, then decreases for $h>h_d$ ($h_d$ is the critical thickness of the misfit dislocation appearance).
- For the tensile strain ($U_m$>0) electret-like phase can transforms into $FE_{aa}$ then into $FE_c$ phases in dependence on the film thickness. The critical thickness of this transition $h_{cr}$ rapidly increases with $U_m$ increase.

• Only for the free standing film ($U_m = 0$) pyroelectric coefficient and static permittivity have divergency at critical size and critical temperature for the ferroelectric to paraelectric phase transition (at $U_m \neq 0$ these properties could have only maximums).

We summarized polar and dielectric properties for all the considered cases ($h<h_{cr}$, $h>h_{cr}$, $U_m>0$, $U_m=0$, $U_m<0$) in the Table 1.

**Table 1. Film properties schematic behavior at $E_0 = 0$ (except $P_Z(E_0)$) for different mismatch strain $U_m$ ($E_m \sim U_m$) at thickness $h < h_{cr}$ and $h > h_{cr}$.**

| Film properties | $h<h_{cr}(T, U_m)$ (see Eqs.(17), (34)) | | | $h>h_{cr}(T, U_m)$ (see Eqs.(17), (34)) | | |
|---|---|---|---|---|---|---|
| | $U_m>0$ | $U_m=0$ | $U_m<0$ | $U_m>0$ | $U_m=0$ | $U_m<0$ |
| $\varepsilon_{ZZ}(T)$ at $E_0=0$ | monotonically decreases at $T$ increasing, $\lvert U_m \rvert$ increasing or $h$ decreasing 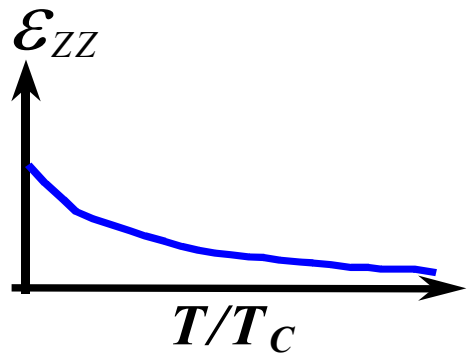  | | | maximum at $T=T_{cr}(h) < T_C$ 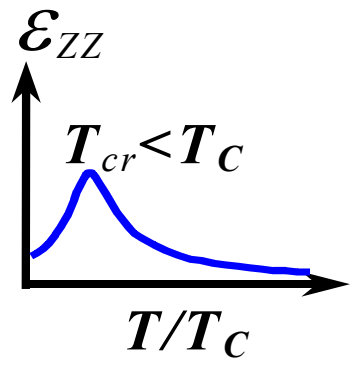  | diverges at $T=T_{cr}(h)$ 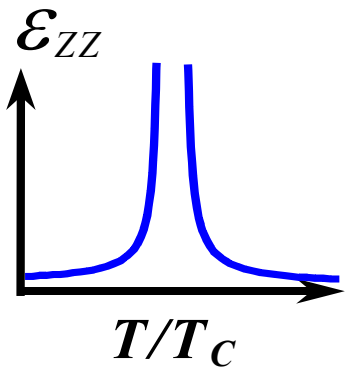  | maximum at $T=T_{cr}(h) > T_C$ 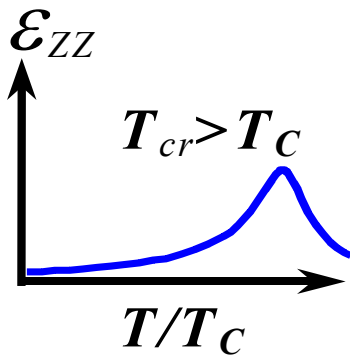  |
| $\Pi_Z(T)$ at $E_0=0$ | monotonically decreases at $T$ increasing, $\lvert U_m \rvert$ increasing or $h$ decreasing | | | negative; maximum at $T=T_{cr}(h) < T_C$ 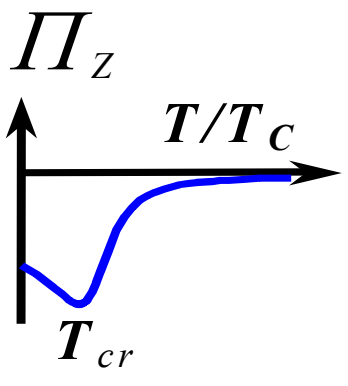  | absent at $T>T_{cr}(h)$; at $T_{cr}(h)$ diverges 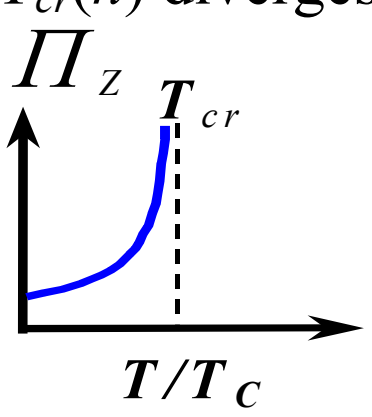  | positive; maximum at $T=T_{cr}(h) > T_C$ 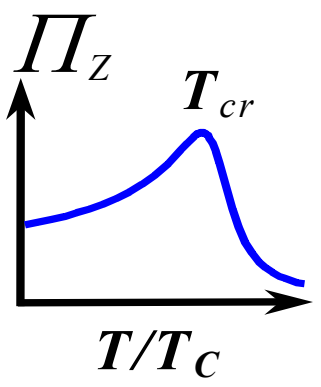  |
| | negative 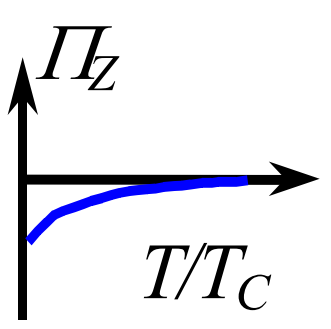  | absent | positive 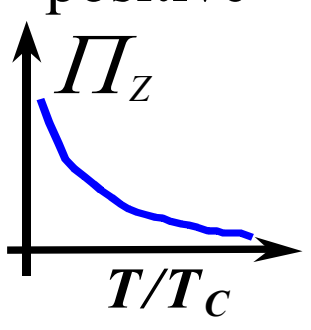  | | | |
| $P_Z(E_0)$, Shift ~ $E_m$ | left-shifted 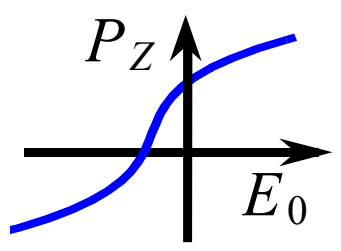  $E_m<0$ | symmetric 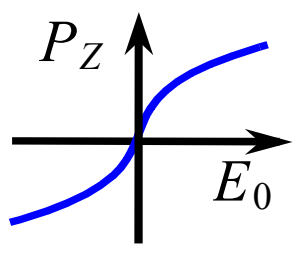  $E_m=0$ | right-shifted 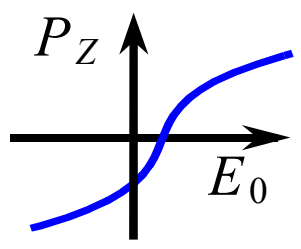  $E_m>0$ | left-shifted 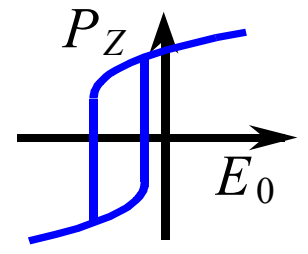  $E_m<0$ | symmetrical 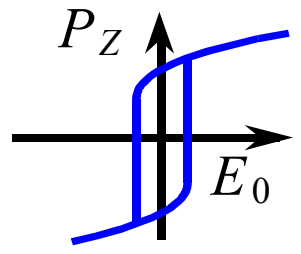  $E_m=0$ | right-shifted 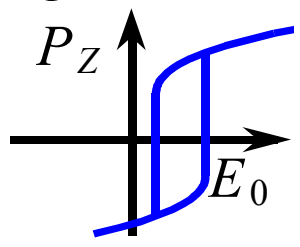  $E_m>0$ |
| $a/c$-ratio at $E_0=0$ | $a \neq c$ piezo-effect | $a=c$ electro-striction | $a \neq c$ piezo-effect | $a \neq c$ piezo-effect | $a \neq c$ piezo-effect | $a \neq c$ piezo-effect |
| Comment | For perovskites on metal-oxide substrates $h_{cr}(U_m<0) < h_{cr}(U_m=0) << h_{cr}(U_m>0)$.<br>The shift of $P_Z(E_0)$ is proportional to $E_m$.<br>For $h > h_{cr}$ maximum smears and flattens with $\lvert U_m \rvert$ increasing or $h$ decreasing | | | | | |

One can see that only in freestanding film ($U_m = 0$) cubic symmetry exists, while at $U_m \neq 0$ there are phases without inversion center and so with piezo- and pyroelectric effects, originated from built in internal field. Electret like state with pyroelectric effect and non-reversible polarization was forecasted for the thin films with $h<h_{cr}$. Conventional ferroelectric state with switchable polarization, permittivity and pyroelectric coefficient temperature maximum was obtained for the films with thickness $h>h_{cr}$. The main condition for the self-polarization appearance was shown to be the inequality $E_m(h) > E_{0c}(h)$ where $E_{0c}(h)$ is the thermodynamic coercive field.

In general case the critical thickness value can be in the region from several to hundreds of lattice constants in dependence on misfit strain, extrapolation length and other materials parameters.

When discussing the question about existence of the thickness induced phase transition for the ultrathin ferroelectric films one has to consider possible appearance of the interface layer consisting of film-substrate material solid solution and try not to misconstrue electret state as true ferroelectric phase. The new experimental investigations aimed to reveal the electret state in ultra-thin films are extremely desirable.

**APPENDIX A**

Linearized solutions of Eq.(10) for paraelectric polarization $P$ and averaged susceptibility $\chi$ have the form:

$$P_Z(z)=\frac{E_Z-2\pi\overline{\varphi(z)})P_m}{\alpha_Z+4\pi\overline{\varphi(z)}}\left[1-\varphi(z)\right]-\frac{P_m}{2}\left[\varphi(z)-\xi(z)\right],\qquad P_{X,Y}(z)=\frac{E_{X,Y}}{\alpha_X}\left(1-\phi(z)\right),$$
$$\overline{\chi_Z(z)}=\frac{1-\overline{\varphi(z)}}{\alpha_Z+4\pi\overline{\varphi(z)}},\qquad \overline{\chi_{X,Y}(z)}=\frac{1-\overline{\phi(z)}}{\alpha_X}. \tag{A.1}$$

Note, that renormalized coefficients near the second power of polarization components are generalized susceptibilities, namely $A_Z\equiv\frac{1}{\overline{\chi_Z(z)}}=\left(\frac{\alpha_Z+4\pi\overline{\varphi(z)}}{1-\overline{\varphi(z)}}\right)$ and $A_X\equiv\frac{1}{\overline{\chi_{X,Y}(z)}}=\frac{\alpha_X}{1-\overline{\varphi(z)}}$. Thus let us find the trial functions in the form, that gives these expressions for $A_{Z,X}$ after averaging over film thickness, namely:

$$P_Z(z)=P_{VZ}\frac{1-\varphi(z)}{1-\overline{\varphi(z)}}-\frac{P_m}{2}\left[\varphi(z)-\xi(z)\right],\qquad P_{X,Y}(z)=P_{V\,X,Y}\frac{1-\phi(z)}{1-\overline{\phi(z)}}. \tag{A.2}$$

The amplitudes $P_{VX,Y,Z}$ are variational parameters. In accordance with (12a) one obtains, that:

$$\overline{\varphi(z)}=\frac{1}{l}\int_{-l/2}^{l/2}dz\frac{ch(z/l_Z)}{ch(l/2l_Z)+(\lambda_Z/l_Z)sh(l/2l_Z)}=\frac{2l_Z}{l}\cdot\frac{sh(l/2l_Z)}{ch(l/2l_Z)+(\lambda_Z/l_Z)sh(l/2l_Z)}=\psi(h) \tag{A.3}$$

Hereinafter we used designations $h=\frac{l}{2\,l_Z}$, $\Lambda_Z=\frac{\lambda_Z}{l_Z}$ and approximation:

$$\psi(h)=\frac{sh(h)}{h\left(ch(h)+\Lambda_Z\,sh(h)\right)}\approx\begin{cases}\dfrac{1}{h(1+\Lambda_Z)}, & h>>1\\[2mm] \dfrac{1}{1+\Lambda_Z h}, & h<<1\end{cases} \tag{A.4}$$

At $\alpha_X>0$ we obtain from (12b) that

$$\overline{\phi(z)}=\frac{1}{l}\int_{-l/2}^{l/2}dz\frac{ch(z/l_X)}{ch(l/2l_X)+(\lambda_X/l_X)sh(l/2l_X)}=\frac{2l_X}{l}\cdot\frac{sh(l/2l_X)}{ch(l/2l_X)+(\lambda_X/l_X)sh(l/2l_X)}. \tag{A.5a}$$

At $\alpha_X<0$ we obtain from (11b) that

$$\overline{\phi(z)}=\frac{1}{l}\int_{-l/2}^{l/2}dz\frac{\cos(z/l_X)}{\cos(l/2l_X)-(\lambda_X/l_X)\sin(l/2l_X)}=\frac{2l_X}{l}\cdot\frac{\sin(l/2l_X)}{\cos(l/2l_X)-(\lambda_X/l_X)\sin(l/2l_X)} \tag{A.5b}$$

Taking into account that $|U_m|<<1$, in accordance to Eq.(9): $U_m^*(T,h)=U_m(T)$ at $h\le h_d$ and $U_m^*(T,h)=\frac{U_m(T)h_d}{h\left(1-U_m(T)(1-h_d/h)\right)}\approx\frac{U_m(T)h_d}{h}$ at $h>h_d$ ($h_d=l_d/2l_Z\sim 1/|U_m|$). For the **second order phase transitions** we obtained from (4) that:

$$A_Z=\frac{\alpha_Z+4\pi\overline{\varphi(z)}}{1-\overline{\varphi(z)}}+\frac{3\beta_Z}{4}P_m^2\overline{\left(\frac{1-\varphi(z)}{1-\overline{\varphi(z)}}\right)^2\left(\varphi^2(z)+\xi^2(z)\right)}\approx$$
$$\approx\frac{\alpha_T}{1-\psi(h)}\left(T-T_C-\frac{2Q_{12}U_m^*}{\alpha_T(S_{11}+S_{12})}+\frac{4\pi}{\alpha_T}\cdot\psi(h)\right)\approx \tag{A.6}$$
$$\approx\begin{cases}\alpha_T\left[T-T_C\left(1+\dfrac{2Q_{12}U_m}{\alpha_T T_C(S_{11}+S_{12})}-\dfrac{4\pi}{(1+\Lambda_Z)\alpha_T T_C}\cdot\psi(h)\right)\right], & h\le h_d\\[3mm] \alpha_T\left[T-T_C\left(1-\left(\dfrac{4\pi}{\alpha_T T_C}\psi(h)-\dfrac{2Q_{12}U_m h_d}{\alpha_T T_C(S_{11}+S_{12})}\dfrac{1}{h}\right)\right)\right], & h>h_d\end{cases}$$

$$E_m = 2\pi\overline{\varphi(z)}\,P_m - \frac{\beta_Z}{8}P_m^3\overline{\left(\varphi^3(z)+3\varphi(z)\,\xi^2(z)\right)\left(\frac{1-\varphi(z)}{1-\overline{\varphi(z)}}\right)} \approx \frac{2\pi U_m}{d_{13}}\psi(h) \qquad \text{(A.7)}$$

In order to estimate the misfit-induced field $E_m$ for $U_m = (0.005 \div -0.005)$, $\Lambda_Z = 50$, $h \sim 10$ we use material coefficients for 50/50 PZT films: $S_{11} = 10.5\cdot 10^{-12}\, m^2/N$, $S_{12} = -3.7\cdot 10^{-12}\, m^2/N$, $Q_{11} = 0.0966\, m^4/C^2$, $Q_{12} = -0.0460\, m^4/C^2$, $d_{31} = -15\cdot 10^{-3}\, Vm/N$ [27]. In SI units one obtains that $E_m = \frac{U_m}{2d_{13}\varepsilon_0}\psi(h) \approx \frac{U_m}{h(1+\Lambda_Z)}3.8\cdot 10^7\frac{kV}{cm} \sim 370\frac{kV}{cm}$, which is compatible with thermodynamic coercive field bulk value $E_{0C} \approx 360\ kV/cm$ calculated at $T = 300K$ (see Fig.5).

For $A_X$ we derived the [1/1]-Pade approximation [35] near the critical thickness $l_{cr}^X = 2l_X arctg(l_X/\lambda_X)$. Firstly we transformed the exact expression:

$$A_X = \alpha_X \frac{1}{1-\frac{2l_X}{l}\cdot\frac{tg(l/2l_X)}{1-(\lambda_X/l_X)tg(l/2l_X)}} = \alpha_X \frac{1-(\lambda_X/l_X)tg(l/2l_X)}{1-(\lambda_X/l_X)tg(l/2l_X)-\frac{2l_X}{l}\cdot tg(l/2l_X)} =$$

$$= \alpha_X \frac{\left(tg(l_{cr}^X/2l_X)-tg(l/2l_X)\right)\frac{\lambda_X}{l_X}}{\left(tg(l_{cr}^X/2l_X)-tg(l/2l_X)\right)\frac{\lambda_X}{l_X}-\frac{2l_X}{l}tg(l/2l_X)}$$

After conventional procedure [35] the [1/1]-Pade approximation over $1/l$ powers has the form:

$$A_X \approx \begin{cases} \alpha_X\left(1-\frac{l_{cr}^X}{l}\right), & l_{cr}^X = 2l_X arctg\left(\frac{l_X}{\lambda_X}\right) \approx \frac{2l_X^2}{\lambda_X}, \quad \left(\frac{\lambda_X}{l_X}\right)^2 >> 1 \\ \alpha_X\left(1-\left(\frac{l_{cr}^X}{l}\right)^2\right), & l_{cr}^X = 2l_X arctg\left(\frac{l_X}{\lambda_X}\right) \approx \pi l_X, \quad \left(\frac{\lambda_X}{l_X}\right)^2 << 1 \end{cases} \qquad \text{(A.8a)}$$

Note, that for the most of cases $(\lambda_X/l_X)^2 = \alpha_X\lambda_X^2/\delta_X \le 10^{-2}\cdot 10^2 nm^2/10^{-15}cm^2 \sim 10^{-1}$. Thus, finally using (A.2) after integration of (4) at $(\lambda_X/l_X)^2 << 1$ we derived that:

$$A_X = \frac{\alpha_X}{1-\overline{\phi(z)}} + \frac{\eta_Z}{4}P_m^2\overline{\left(\frac{1-\varphi(z)}{1-\overline{\varphi(z)}}\right)^2\left(\varphi^2(z)+\xi^2(z)\right)} \approx$$

$$\approx \alpha_T\left(T - T_C^X(U_m^*)\right)\left(1-\left(\frac{l_{cr}^X(T,U_m^*)}{l}\right)^2\right) + \frac{\eta_Z}{d_{13}^2}\frac{U_m^2}{4h(1+\Lambda_Z)^2} \qquad \text{(A.8b)}$$

Here the critical thickness is introduced:

$$l_{cr}^X(T,U_m^*) \approx \pi\sqrt{\frac{\delta_X}{\alpha_T\left(T_C^X(U_m^*)-T\right)}}, \qquad T_C^X(U_m^*) = T_C + \frac{(Q_{11}+Q_{12})U_m^*}{\alpha_T(S_{11}+S_{12})}. \qquad \text{(A.9)}$$

Approximate expressions for higher coefficients could be obtained similarly, but much more cumbersome, namely for $h(1+\Lambda_Z) >> 1$ and neglecting terms with $U_m$ higher powers we derived that:

$$B_Z = \beta_Z\overline{\left(\frac{1-\varphi(z)}{1-\overline{\varphi(z)}}\right)^4} \approx \beta_Z, \qquad C_Z = \gamma_Z\overline{\left(\frac{1-\varphi(z)}{1-\overline{\varphi(z)}}\right)^6} \approx \gamma_Z, \qquad \text{(A.10)}$$

$$D_m = -\frac{3\beta_Z}{2}P_m\overline{\left(\frac{1-\varphi(z)}{1-\overline{\varphi(z)}}\right)^3\varphi(z)} \approx -\frac{3\beta_Z}{2d_{13}}\frac{U_m}{h(1+\Lambda_Z)}, \qquad \text{(A.11)}$$

$$H_m = -\frac{5\gamma_Z}{2} P_m \overline{\left(\frac{1-\varphi(z)}{1-\overline{\overline{\varphi(z)}}}\right)^5 \varphi(z)} \approx -\frac{5\gamma_Z U_m}{2d_{13}} \cdot \frac{1}{h(1+\Lambda_Z)}, \qquad B_X = \beta_X \overline{\left(\frac{1-\phi(z)}{1-\overline{\overline{\phi(z)}}}\right)^4} \approx \beta_X \tag{A.12}$$

$$C_X = \gamma_X \overline{\left(\frac{1-\phi(z)}{1-\overline{\overline{\phi(z)}}}\right)^6} \approx \gamma_X,\ C_{XY} = \eta_X \overline{\left(\frac{1-\phi(z)}{1-\overline{\overline{\phi(z)}}}\right)^4} \approx \eta_X,\ F_{XZ} = \eta_Z \overline{\left(\frac{1-\phi(z)}{1-\overline{\overline{\phi(z)}}}\right)^2 \left(\frac{1-\varphi(z)}{1-\overline{\overline{\varphi(z)}}}\right)^2} \approx \eta_Z \tag{A.13}$$

$$K_m = \eta_Z P_m \overline{\left(\frac{1-\phi(z)}{1-\overline{\overline{\phi(z)}}}\right)^2 \left(\frac{1-\varphi(z)}{1-\overline{\overline{\varphi(z)}}}\right) \varphi(z)} \approx \frac{\eta_Z}{d_{13}} \frac{U_m}{2} \frac{1}{h\left(1+\Lambda_Z\right)}. \tag{A.14}$$

$$L_m = \frac{\gamma_{XYZ}}{2} P_m \overline{\left(\frac{1-\phi(z)}{1-\overline{\overline{\phi(z)}}}\right)^4 \left(\frac{1-\varphi(z)}{1-\overline{\overline{\varphi(z)}}}\right) \varphi(z)} \approx \frac{\gamma_{XYZ}}{2} \frac{U_m}{d_{13}} \frac{1}{h(1+\Lambda_Z)}, \tag{A.15}$$

$$F_m = \frac{\eta_{XYZ}}{2} P_m \overline{\left(\frac{1-\phi(z)}{1-\overline{\overline{\phi(z)}}}\right)^4 \left(\frac{1-\varphi(z)}{1-\overline{\overline{\varphi(z)}}}\right) \varphi(z)} \approx \frac{\eta_{XYZ}}{2} \frac{U_m}{d_{13}} \frac{1}{h(1+\Lambda_Z)} \tag{A.16}$$

$$C_{XYZ} = \gamma_{XYZ} \overline{\left(\frac{1-\phi(z)}{1-\overline{\overline{\phi(z)}}}\right)^4 \left(\frac{1-\varphi(z)}{1-\overline{\overline{\varphi(z)}}}\right)^2} \approx \gamma_{XYZ}, \qquad F_{XXX} = \eta_{XYZ} \overline{\left(\frac{1-\phi(z)}{1-\overline{\overline{\phi(z)}}}\right)^6} \approx \eta_{XYZ} \tag{A.17}$$

$$F_{XYZ} = \eta_{XYZ} \overline{\left(\frac{1-\phi(z)}{1-\overline{\overline{\phi(z)}}}\right)^2 \left(\frac{1-\varphi(z)}{1-\overline{\overline{\varphi(z)}}}\right)^4} \approx \eta_{XYZ},\ Q_{XYZ} = \eta_{XYZ} \overline{\left(\frac{1-\phi(z)}{1-\overline{\overline{\phi(z)}}}\right)^4 \left(\frac{1-\varphi(z)}{1-\overline{\overline{\varphi(z)}}}\right)^2} \approx \eta_{XYZ} \tag{A.18}$$